\documentclass[11pt]{article}
\pdfoutput=1
\usepackage{jcappub,natbib}
\input{colordvi.tex}

\def\ga{\mathrel{\raise.3ex\hbox{$>$\kern-.75em\lower1ex\hbox{$\sim$}}}}
\def\la{\mathrel{\raise.3ex\hbox{$<$\kern-.75em\lower1ex\hbox{$\sim$}}}}

\def\NL{\text{\tiny NL}}

\def\GW{\text{\tiny GW}}
\renewcommand{\la}{\left \langle}

\usepackage{hyperref}

\title{Title} 

\abstract{An observable stochastic background of gravitational waves is generated whenever primordial black holes are created in the early universe thanks to a small-scale enhancement of the curvature perturbation. We calculate the anisotropies and  non-Gaussianity of such stochastic gravitational waves background which receive two contributions, the first at formation time and the second due to propagation effects. We conclude that  a sizeable  magnitude of anisotropy and non-Gaussianity in the gravitational waves would suggest that primordial black holes may not comply the totality of the dark matter. }

\title{Title}

\author[a,b,c]{Nicola Bartolo\,}
\author[a,b]{, Daniele Bertacca\,}
\author[a,b,c,d]{, Sabino Matarrese\,}
\author[a,b]{Marco Peloso\,}
\author[b]{, Angelo Ricciardone\,}
\author[e,f]{, Antonio Riotto\,}
\author[g]{, Gianmassimo Tasinato\,}

\title{ Characterizing the Cosmological Gravitational Wave Background:\\
Anisotropies and non-Gaussianity  }

\affiliation[a]{Dipartimento di Fisica e Astronomia ``Galileo Galilei'', Universit\'a di Padova, 35131 Padova, Italy}
\affiliation[b]{INFN, Sezione di Padova, 35131 Padova, Italy}
\affiliation[c]{INAF - Osservatorio Astronomico di Padova, I-35122 Padova, Italy}
\affiliation[d]{Gran Sasso Science Institute, I-67100 L'Aquila, Italy}
\affiliation[e]{Department of Theoretical Physics and Center for Astroparticle Physics (CAP), CH-1211 Geneva 4, Switzerland}
\affiliation[f]{CERN,Theoretical Physics Department, Geneva, Switzerland}
\affiliation[g]{Department of Physics, Swansea University, Swansea, SA2 8PP, UK}

\abstract{A future detection of the Stochastic  Gravitational Wave Background  (SGWB) with GW experiments  is expected to open a new window on  early universe cosmology and on the astrophysics of compact objects. In this paper we  study  SGWB anisotropies,  that can offer new tools  to discriminate between different sources of GWs. 
In particular, the cosmological SGWB inherits its anisotropies both (i) at its production and (ii) during its propagation through our perturbed universe. Concerning (i), we show that it typically leads to anisotropies with order one dependence on frequency. We then compute the effect of (ii) through a Boltzmann approach, including contributions   of both large-scale scalar and tensor linearized perturbations. We  also compute for the first time the three-point function of the SGWB energy density, which can allow one to extract information on GW non-Gaussianity  with interferometers. Finally, we include non-linear effects associated with long wavelength scalar fluctuations, and compute the   squeezed limit of the 3-point function for the SGWB density contrast. Such limit satisfies a consistency relation, conceptually similar to what found in the literature for the case of  CMB perturbations.}

\begin{document}
\maketitle
\flushbottom

\section{Introduction}
\label{sec:intro}

The current ground based interferometers are close to reach the expected sensitivity to detect the Stochastic Gravitational Wave Background (SGWB) from unresolved astrophysical sources \cite{LIGOScientific:2019vic}. Future space-based (such as LISA \cite{Audley:2017drz} and DECIGO \cite{Kawamura:2006up}) and earth-based (like Einstein Telescope \cite{Sathyaprakash:2011bh, Maggiore:2019uih} and Cosmic Explorer \cite{Evans:2016mbw}) interferometers have the potential to detect the SGWB of cosmological origin (see  \cite{Maggiore:1999vm,Guzzetti:2016mkm,Bartolo:2016ami,Caprini:2018mtu} for reviews of possible cosmological sources). It is likely that a detection of a cosmological SGWB background will require the ability to discriminate it against the astrophysical signal. 
Astrophysical GW background (AGWB) arises from the superposition of the signals emitted by a large population of unresolved sources that are  mainly dominated by two types of events: (i) the periodic long lived sources (e.g.  the early inspiraling phase of binary systems) where the frequency is expected to evolve very slowly compared to the observation time; (ii) the short-lived burst sources, e.g. core collapse to neutron stars or black holes, oscillation modes, r-mode instabilities in rotating neutron stars, magnetars and super-radiant instabilities (for example, see \cite{Regimbau:2011rp,Romano:2016dpx}).  Several techniques have been developed to distinguish among the various backgrounds.
The most obvious tool for this component separation  is the frequency dependence  \cite{Caprini:2019pxz}, as several cosmological mechanisms are peaked at some given characteristic scale. 
However, future detectors  will allow for a better angular resolution of anisotropies of the astrophysical background. Therefore, another tool could be the directionality dependence of the SGWB \cite{Alba:2015cms,Contaldi:2016koz, Cusin:2017fwz, Jenkins:2018lvb, Cusin:2018avf,Bertacca:2019fnt} and, as we explore here, its statistics.

In this work, we discuss  graviton propagation through a Boltzmann approach \cite{Contaldi:2016koz} as it is typically done for the CMB. Specifically, we construct and evolve the equation for the distribution  $f$ of  gravitons in a FLRW background,  plus first order scalar and tensor perturbations (we also consider how non-linear effects for the specific case of squeezed non-Gaussianity, as we discuss at the end of this Introduction). At the unperturbed level, following the isotropy and homogeneity of the background, the distribution depends only on time and on the GW frequency $p/2 \pi$ (where $\vec{p}$ is the physical momentum of the gravitons) %, and only 
 through the combination $q \equiv p \, a$, where $a$ is the scale factor of the universe. Namely, the gravitons freely propagate, and their physical momentum redshifts during the propagation. This property is shared by any free massless particles, and, in particular, also by the CMB photons. On the other hand, differently from the photon distribution, the initial population of gravitons is not expected to be thermal (as we have in mind production mechanisms, such as inflation \cite{Barnaby:2010vf,Cook:2011hg}, phase transitions \cite{Geller:2018mwu}, or enhanced density perturbations leading to primordial black holes (PBH) \cite{Bartolo:2018evs,Bartolo:2018rku,Bartolo:2019zvb}, which occur at energies well below the Planck scale) which leaves in the distribution a sort of ``memory'' of the initial state. As we show, the fact that the spectrum is non thermal generically results in angular anisotropies that have an order one dependence on the GW frequency. This is in contrast with the CMB case, for which this dependence only arises at second order in perturbation theory. 

This initial state will in general be anisotropic, as no mechanism of GW production can be perfectly homogeneous. Additional anisotropies are induced by the GW propagation in the perturbed universe. As we are interested in large scale, we work in a regime of a large hierarchy $q \gg k$ between the GW (comoving) momentum $q$ and the (comoving) momentum $k$ of the large scale perturbations. We confirm that in the angular power spectrum, the  Sachs Wolfe (SW) effect is dominating on large scales also for gravitons, while the Integrated Sachs-Wolfe (ISW) contribution is subdominant.  

We employ this approach to study the non-Gaussianity of the SGWB energy density. Although we are not aware of any dedicated analysis in this sense, it is reasonable to expect that the SGWB produced by incoherent astrophysical sources is Gaussian, due to the central limit theorem. In light of this fact, a measurement of non-Gaussianity would be a signal of large scale coherency, that would likely point to a cosmological origin of the signal. Previous works showed that inflation can result in a sizeable an nonvanishing $3-$point function $\left\langle h^3 \right\rangle$ for the graviton wave function, but that this is generically non observable in interferometers \cite{Bartolo:2018evs,Bartolo:2018rku}, due to the decoherence of the phase  the GW wave-function $h$ induced by the GW propagation, and due to the finite duration of the measurement (see \cite{Powell:2019kid} for a possible exception to this conclusion, occurring for a very specific shape of the bispectrum). Since the phase does not affect the GW energy density, we argue that the energy density is a much better variable to study the statistics of the SGWB. Also in this case, the of non-Gaussianity can be induced both by the production mechanism and the propagation. As an example of the former, in ref. \cite{Bartolo:2019zvb} we recently computed the $3-$point function of the SGWB energy density that arises in presence of non-Gaussianity of the scalar perturbations of the local shape (in presence of this non-Gaussianity, a long-scale mode of momentum $k$ can modulate the power of the short-scalr scalar perturbations  that are responsible for the PBH formation). Here we study the  $3-$point function induced by the GW propagation. This is also proportional to the non-Gaussianity of the scalar perturbations. In this sense, the SGWB can be used as a novel probe (beyond the CMB and the LSS) of the non-Gaussianity of the scalar perturbations. 

Although in most of this work we limit our attention to linearized fluctuations, in Section  \ref{sec-sqz}  we consider non-linear effects induced by long-wavelength scalar perturbations, which modulate correlation functions involving short-wavelength modes. We make use of a powerful method first introduced by Weinberg in \cite{Weinberg:2003sw}, which focusses on adiabatic systems, and identifies  the effects of long modes with an appropriate coordinate transformation. Applying this method to our set-up, we compute how non-linearities induce a non-vanishing   squeezed limit of the 3-point function for the SGWB density contrast. We determine how such squeezed limit depends on the scale-dependence of the spectrum of primordial scalar fluctuations; on the momentum dependence of the background SGWB distribution; and on  the time, scale, and direction dependence of the scalar transfer functions connecting primordial to late-time adiabatic scalar fluctuations.

The paper is organized as follows. In Section \ref{sec: BoltzmannGW} we present the computation and the formal solution of the 
Boltzmann equation for GW propagation. In Section \ref{sec:harmsol} we decompose the formal solution in spherical harmonics,  paralleling a  treatment that is familiar in the study of CMB perturbations. In Section \ref{sec: GWcorrelator} we compute the angular power spectrum and bispectrum of the SGWB perturbations. In Section \ref{sec: spectrald} we review one physical mechanism that can result in a sizeable cosmological SGWB with some degree of anisotropy. In Section \ref{sec-sqz} we study non-linear effects on the squeezed bispectrum.  These results are discussed and summarized in Section \ref{sec: conclusion}. The paper is concluded by three appendices.  Appendix \ref{app:tensor-sourced} contains the details of the computation of the anisotropies due to the large-scale tensor perturbations.  Appendix \ref{app:tensor2B} provides some intermediate steps on the computation of the GW bispectrum induced by tensor modes. Finally, Appendix \ref{app: comparison} presents an immediate connection between our formal solutions and the CMB results obtained in the case of initial thermal state. 

Part of the results contained  in the present work were also summarized in the Rapid Communication \cite{Bartolo:2019oiq}.

\section{Boltzmann equation for gravitational waves}
\label{sec: BoltzmannGW}

We consider first order perturbations around a Friedmann-Lemaitre-Robertson-Walker (FLRW) background in the Poisson gauge 
\begin{equation}
\label{metric}
ds^2=a^2(\eta)\left[
-e^{2\Phi} d\eta^2+(e^{-2\Psi}\delta_{ij}+ \chi_{ij}) 
dx^i dx^j\right]\, ,
\end{equation}
where $a(\eta)$ is the scale factor as a function of the conformal time 
$\eta$. $\Phi$ and $\Psi$ are scalar perturbations while  $\chi_{ij}$ 
represent the transverse-traceless (TT) tensor perturbations. We neglect linear 
vector modes since they are not produced at first order in standard 
mechanisms for the generation of cosmological perturbations 
(as scalar field inflation), and 
 we consider tensor modes at linearised order.

Given the statistical nature of the GW we can define a distribution function of gravitons as $f = f ( x^{\mu}, p^{\mu})$, which is function of their position $x^{\mu}$ and momentum $p^{\mu} = d x^{\mu}/d \lambda$, where $\lambda$ is an affine parameter along the GW trajectory.
As we will see, observables as number density, spectral energy density, and flux (directions) can be derived from the distribution function. The graviton distribution function obeys the Boltzmann equation
\begin{equation}
\label{Boltzgeneric}
%\frac{df}{d\eta}= {\overline C}[f] + \, ,
\mathcal{L}[f] = \mathcal{C}[f(\lambda)] + \mathcal{I}[f(\lambda)]\,,
\end{equation}  
where  $\mathcal{L}\equiv d/ d\lambda$ is the Liouville term, while $\mathcal{C}$ and $ \mathcal{I}$ account, respectively, for the collision of  GWs along their path, and for their emissivity from cosmological and astrophysical sources \cite{Contaldi:2016koz}. The collision among GWs affects  the distribution  at higher orders  (in an expansion series in the gravitational strength $1/M_P$, where $M_P$ is the Planck mass) with respect to the ones we are considering, and they can be disregarded in our analysis  (see \cite{Bartolo:2018igk} and references  therein for a discussion of collisional effects involving gravitons).  The emissivity can be due to astrophysical processes (such as black hole merging) in the relatively late universe, as well as cosmological processes, such as inflation or phase transitions. In this work we are only interested in the stochastic GW background of cosmological origin, so we treat the emissivity term as an initial condition on the GW distribution. This leads us to study the free Boltzmann equation, $d f/d \eta = 0$ in the perturbed universe
\begin{equation}
\label{Dfc}
\frac{d f}{d \eta} = \frac{\partial f}{\partial \eta}+
\frac{\partial f}{\partial x^i} \frac{d x^i}{d \eta}+
\frac{\partial f}{\partial q} \frac{d q}{d \eta}+
\frac{\partial f}{\partial n^i} \frac{d n^i} {d \eta} =0\, , 
\end{equation}
where ${\hat n} \equiv {\hat p}$ is the GW  direction of motion, and where we have used the comoving momentum $q \equiv \vert \vec{p} \vert a$ (as opposed to the physical one, used in  \cite{Dodelson:2003ft, Contaldi:2016koz}). This  simplifies the equations by factorizing out the universe expansion. The first two terms in (\ref{Dfc}) encode free streaming, that is the propagation of perturbations on all scales.  At higher order this term also includes gravitational time delay effects. The third term causes the red-shifting of gravitons, including the Sachs-Wolfe (SW), integrated Sachs-Wolfe (ISW) and Rees-Sciama (RS) effects. The fourth term vanishes to first order,  and describes the effect of gravitational lensing. We shall refer to these terms as the free-streaming, redshift and lensing terms, respectively, as customarily done in  CMB physics.  

Keeping only the terms up to first order in the perturbations, Eq. (\ref{Dfc}) gives 
\begin{equation}
\frac{\partial f}{\partial \eta}+
n^i \, \frac{\partial f}{\partial x^i} +
\left[  \frac{\partial \Psi}{\partial \eta} - n^i \, \frac{\partial \Phi}{\partial x^i} + \frac{1}{2}  n_i n_j \frac{\partial  \chi_{ij} }{\partial \eta} \right] q \,   \frac{\partial f}{\partial q} = 0 \,,
\label{Dfc2}
\end{equation}
where we have followed the standard procedure developed for the CMB in  \cite{Dodelson:2003ft,Bartolo:2006fj}.
The distribution function $f$ can be expanded as 
\begin{equation}
f \left( \eta ,\, x^i ,\, q ,\, n^i \right) = {\bar f} \left( q \right) +  f^{(1)}  \left( \eta ,\, x^i ,\, q ,\, n^i \right)+....  
\equiv  {\bar f} \left( q \right)  - q \, \frac{\partial {\bar f}}{\partial q} \, \Gamma \left( \eta ,\, x^i ,\, q ,\, n^i \right) +.... \;, 
\label{Gamma-def}
\end{equation} 
where the dominant, homogeneous and isotropic contribution ${\bar f}(q)$ solves the zeroth order Boltzmann equation. The function $ f^{(1)}  ( \eta ,\, x^i ,\, q ,\, n^i )$ is the solution of the first order equation, and the ellipses denote the higher order solutions in a 
 perturbative expansion. In  this expression we have parameterized  the first order solution in terms of the function $\Gamma$, so to simplify the first order Boltzmann equation  \cite{Contaldi:2016koz}. For a  thermal distribution with temperature $T$, one finds $\Gamma = \delta T/T$. This is particularly the case 
for the CMB, for which, due to the thermalization, the temperature anisotropies are frequency-independent up to second order in the  perturbations. For gravitons, as we already mentioned, the collisional term is extremely small, and, for a generic production mechanism, $\Gamma$ generically retains an order one dependence on frequency (as we show below, also for the GW case the propagation effects induce frequency-independent perturbations at linear order). 

The zeroth order homogeneous Boltzmann equation simply reads $\partial \bar{f}/ \partial \eta=0$, and it is solved by any distribution that is function only of the comoving momentum $q$, namely $f = {\bar f} \left( q \right)$. In our approach this solution is simply given as the homogeneous part of the initial condition. As a consequence, the physical momentum of the individual gravitons redshifts proportionally to $1/a$, and the physical graviton number density $n \propto \int d^3 p \, {\bar f} ( q )$ is diluted as $a^{-3}$ as the universe expands. This is also the case for  CMB photons, whose distribution function $ {\bar f}_{CMB} = (e^{p/T}-1)^{-1}$ is only controlled by the ratio $p/T\propto a\,p = q$, where $T$ is the temperature of the CMB bath. We see that these rescalings with $a$ are a  consequence of the free particle propagation in the expanding FLRW background, and they do not rely on the distribution being thermal.  

As anticipated, from the graviton distribution function, evaluated at the present time $\eta_0$, we can compute the SGWB energy density 
\begin{eqnarray}
\rho_\GW \left( \eta_0 ,\, \vec{x} \right) &=& \frac{1}{a_0^4}  \int d^3 q \, q \, f \left( \eta_0 ,\, \vec{x} ,\, q ,\, {\hat n} \right) 
\equiv \rho_{\rm crit,0} \, \int d \ln q \; \Omega_\GW \left(  \vec{x} ,\, q  \right) \,,
\label{rho-GW}
\end{eqnarray}
where we have introduced the spectral energy density  $\Omega_\GW$ and the critical density $ \rho_{\rm crit}= 3 H^2 M_p^2$. Here $H \equiv (1/a^2) \, d a/d \eta$ is the Hubble rate. Following standard conventions, the suffix $0$ denotes a quantity evaluated at the present time. 

Contrary to most of the studies of the SGWB, that assume a homogeneous $\Omega_\GW$, in our case the GW energy density depends on space. We denote the homogeneous component of  $\Omega_\GW$ as 
\begin{equation}
\bar{\Omega}_\GW \left( q \right) \equiv \frac{4 \pi}{\rho_{\rm crit,0}} \, \left( \frac{q}{a_0} \right)^4 {\bar f} \left( q \right) \;,  
\end{equation} 
For the full  spectral energy density, we define 
\begin{equation}
\Omega_\GW \equiv \frac{1}{4 \pi} \,  \int d^2 {\hat n} \; \omega_\GW (  \vec{x} ,\, q ,\, {\hat n} ) \;, 
\end{equation}
and we introduce the SGWB density contrast 
\begin{equation}
\delta_\GW \equiv  \frac{\omega_\GW (  \vec{x} ,\, q ,\, {\hat n} ) - {\bar \Omega}_\GW ( q ) }{\bar \Omega_\GW ( q ) } 
 = \left[ 4 -  \frac{\partial \ln \, {\bar \Omega}_\GW \left(  q \right)}{\partial \ln \, q}  \right] \, \Gamma \left( \eta_0 ,\, \vec{x} ,\, q ,\, {\hat n} \right) \,. 
\label{delta-Gamma}
\end{equation}

In terms of the function $\Gamma$, the first order Boltzmann equation reads \cite{Contaldi:2016koz} 
\begin{equation}
\label{Boltfirstgamma1}
 \frac{\partial \Gamma}{\partial \eta}+
n^i \frac{\partial  \Gamma}{\partial x^i}  = S \left(\eta, x^i, n^i\right)  \, ,
\end{equation}
where $$ S  \left(\eta, x^i, n^i\right)  = \frac{\partial \Psi }{\partial \eta } - n^i  \frac{\partial  \Phi}{\partial x^i} -  \frac{1}{2}n^i n^j  \frac{\partial \chi_{ij} }{\partial \eta } $$  is the source function which includes the physical effects due to cosmological scalar and tensor inhomogeneities.  We note that the source is $q-$independent (thus showing that the anisotropies arising at first order from  propagation effects 
are frequency-independent, as we anticipated). 

To solve this equation, it is convenient to Fourier transform with respect to  spatial coordinates, 
\begin{equation}
\Gamma  \equiv \int \frac{d^3 k}{\left( 2 \pi \right)^3} {\rm e}^{i \vec{k} \cdot \vec{x}} \Gamma \left( \eta ,\, \vec{k} ,\, q ,\, {\hat n} \right) \;, 
\end{equation} 
and analogously for the other variables (we use the same notation for a field and for its Fourier transform, as the context always clarifies which object we are referring to). This leads to 
\begin{equation}
\Gamma'+ i \, k \, \mu\, \Gamma = S (\eta, \vec{k}, {\hat n})  \, ,
\label{Boltfirstgamma1}
\end{equation}
where from now on prime denotes a derivative with respect to conformal time, and where we denote by 
\begin{equation}
\mu \equiv {\hat k} \cdot {\hat n}  \,,
\end{equation}
the cosine of the angle between the Fourier variable $\vec{k}$ and the direction of motion ${\hat n}$ of the GW. In Fourier space the source term reads 
\begin{equation}
S  = \Psi' - i k \, \mu \, \Phi -  \frac{1}{2}n^i n^j  \, \chi_{ij}' \,.
\end{equation}
With this information in mind, Eq. (\ref{Boltfirstgamma1}) is readily integrated to give 
\begin{eqnarray}
\Gamma \left( \eta ,\, \vec{k} ,\, q ,\, {\hat n} \right) &=& {\rm e}^{i k \mu \left( \eta_{\rm in} - \eta \right)} \, \Gamma \left( \eta_{\rm in}  ,\, \vec{k} ,\, q ,\, {\hat n} \right) \nonumber\\ 
&& + \int_{\eta_{\rm in}}^\eta d \eta' \, {\rm e}^{i k \mu \left( \eta' - \eta \right)} \left[ \frac{d \Psi \left( \eta' ,\, \vec{k} \right)}{d \eta' } - i k \mu \Phi \left( \eta' ,\, \vec{k} \right) - \frac{1}{2}   n^i n^j \frac{\partial \chi_{ij} \left( \eta' ,\, \vec{k} \right)}{\partial \eta'} \right] \,. 
\nonumber\\ 
\end{eqnarray} 
We integrate the second term in the second line by parts, and obtain
\begin{eqnarray}
\Gamma \left( \eta ,\, \vec{k} ,\, q ,\, {\hat n} \right) &=& {\rm e}^{i k \mu \left( \eta_{\rm in} - \eta \right)} \, \left[ \Gamma \left( \eta_{\rm in} ,\, \vec{k} ,\,q ,\, {\hat n} \right) + \Phi \left( \eta_{\rm in} ,\, \vec{k} \right) \right] - \Phi \left( \eta ,\, \vec{k} \right) \nonumber\\ 
&& \!\!\!\!\!\!\!\!  \!\!\!\!\!\!\!\!  \!\!\!\!\!\!\!\!  \!\!\!\!\!\!\!\! 
+ \int_{\eta_{\rm in}}^\eta d \eta' \, {\rm e}^{i k \mu \left( \eta' - \eta \right)} \left\{ \frac{d \left[ \Psi \left( \eta' ,\, \vec{k} \right) +  \Phi \left( \eta' ,\, \vec{k} \right) \right]}{d \eta' } - i k \mu \Phi \left( \eta' ,\, \vec{k} \right) - \frac{1}{2}   n^i n^j \frac{\partial {\hat \chi}_{ij} \left( \eta' ,\, \vec{k} \right)}{\partial \eta'} \right\} \,, 
\nonumber\\ 
\end{eqnarray} 
with the last two terms in the first line  being the boundary terms of this integration. In the following section, we decompose the ${\hat n}$-dependence of the solution (representing the arrival direction of the GW on our sky) in spherical harmonics. As we are not interested in the monopole term, we can disregard the $ - \Phi ( \eta ,\, \vec{k} )$ contribution to the solution, and write 
\begin{multline} 
\Gamma \left( \eta ,\, \vec{k} ,\, q ,\, {\hat n} \right) \equiv \int_{\eta_{\rm in}}^\eta d \eta' \, {\rm e}^{i k \mu \left( \eta' - \eta \right)}  \times\Bigg\{ \left[ \Gamma \left( \eta' ,\, \vec{k} ,\, q ,\, {\hat n} \right)  + \Phi \left( \eta' ,\, \vec{k} \right) \right] 
\delta \left( \eta' - \eta_{\rm in} \right)  \phantom{xxxx}\\
+  \frac{\partial \left[ \Psi \left( \eta' ,\, \vec{k} \right) +  \Phi \left( \eta' ,\, \vec{k} \right) \right] }{\partial \eta'} 
-\frac{1}{2}  n^i n^j \frac{\partial {\hat \chi}_{ij} \left( \eta' ,\, \vec{k} \right)}{\partial \eta'} \Bigg\}\,. 
\label{solution}
\end{multline} 
The first  term, which was disregarded in \cite{Contaldi:2016koz},  carries the ``memory'' of the initial conditions. Due to this term, the GW energy density anisotropies are generically dependent on the frequency $q$.  We discuss an example of this fact in Section \ref{sec: spectrald}, where we study the SGWB produced in axion inflation. 

Generally, this term has also a  dependence on ${\hat n}$. This implies that   the solution has a dependence on the direction $\hat n$,  which  is more general than the one arising from the projection of $\vec{k}$ on the line of sight ${\hat n}$. (Indeed, the remaining terms in Eq. (\ref{solution}) depend on ${\hat n}$ only through the $\mu \equiv  {\hat k} \cdot {\hat n} $ combination.   Thanks to this fact, they result in angular correlators that are statistically isotropic (as we show in the next two sections).) On the other hand, the angular dependence present in the first term of (\ref{solution}) could result in statistically anisotropic correlators (specifically, 2-point and 3-point correlators) that have a  more general dependence on the multipoles coefficients $\ell_i$ and $m_i$ than Eqs.  (\ref{Cell-Bell-def}). This would indicate an overall anisotropy of the mechanism responsible for the GW across the entire universe, and, ultimately, a departure from an exact FLRW geometry. While we believe that this can be an interesting topic for future exploration, the present work focuses on the statistically isotropic case, and we assume an initial condition of the form $\Gamma_{\rm in} = \Gamma ( \eta_{\rm in} ,\, \vec{k} ,\, q )$, 
which guarantees such a condition.

\section{Spherical harmonics decomposition}
%Solution for coefficients of spherical harmonics} 
\label{sec:harmsol}

We separate the solution (\ref{solution}) in three terms 
\begin{equation} 
\Gamma \left( \eta ,\, \vec{k} ,\, q ,\, {\hat n} \right) =  \Gamma_I \left( \eta ,\, \vec{k} ,\, q ,\, {\hat n} \right) +  \Gamma_S \left( \eta ,\, \vec{k} ,\,  {\hat n} \right) +  \Gamma_T \left( \eta ,\, \vec{k} ,\,  {\hat n} \right) \;, 
\end{equation}
where $I$, $S$, and $T$ stand for {\it Initial}, {\it Scalar} and {\it Tensor} sourced terms respectively and they are given by
\begin{eqnarray}
\Gamma_I \left( \eta ,\, \vec{k} ,\, q ,\, {\hat n} \right) &=& {\rm e}^{i k \mu \left( \eta_{\rm in} - \eta \right)} \Gamma \left( \eta_{\rm in} ,\, \vec{k} ,\, q  \right) \,,\nonumber\\ 
\Gamma_S \left( \eta ,\, \vec{k} ,\,  {\hat n} \right) &=&  \int_{\eta_{\rm in}}^\eta d \eta' \, {\rm e}^{i k \mu \left( \eta' - \eta \right)} \left[  \Phi \left( \eta' ,\, \vec{k} \right) \delta \left( \eta' - \eta_{\rm in} \right) + 
\frac{\partial \left[ \Psi \left( \eta' ,\, \vec{k} \right) +  \Phi \left( \eta' ,\, \vec{k} \right) \right] }{\partial \eta'} \right] \,,\nonumber\\ 
\Gamma_T \left( \eta ,\, \vec{k} ,\,  {\hat n} \right) &=& - \frac{n^i \, n^j}{2} \,   \int_{\eta_{\rm in}}^\eta d \eta' \, {\rm e}^{i k \mu \left( \eta' - \eta \right)} \, \frac{\partial {\hat \chi}_{ij} \left( \eta' ,\, \vec{k} \right)}{\partial \eta'} \,.
\end{eqnarray} 

Similarly to what is usually done for the CMB, in order to compute the angular power spectrum, in an all-sky analysis  we decompose the  fluctuations using spin-0 or spin-2 spherical harmonics. Since $\Gamma$ is a scalar, we can express it as 
\begin{equation}
\Gamma \left( {\hat n} \right) = \sum_\ell \sum_{m = - \ell}^\ell \Gamma_{\ell m} \, Y_{\ell m} \left( {\hat n} \right) \;\;\;, \; {\rm inverted \; by }  \;\;\Gamma_{\ell m} = \int d^2 n \, \Gamma \left( {\hat n} \right)  \, Y_{\ell m}^* \left( {\hat n} \right) \,,
\end{equation} 
where we recall that $\hat{n}$ is the direction of motion of the GWs.
% and so the direction at which the GWs arrive on our sky.
More specifically:
% for our solution in harmonic space we have 
%
\begin{eqnarray} 
\Gamma_{\ell m} &=& \int d^2 n \,  Y_{\ell m}^* \left( {\hat n} \right) \int \frac{d^3 k}{\left( 2 \pi \right)^3} \, {\rm e}^{i \vec{k} \cdot \vec{x}} 
\left[  \Gamma_I \left( \eta ,\, \vec{k} ,\, q ,\, {\hat n} \right) +  \Gamma_S \left( \eta ,\, \vec{k} ,\,  {\hat n} \right) +  \Gamma_T \left( \eta ,\, \vec{k} ,\,  {\hat n} \right) \right] \nonumber\\ 
&\equiv& \Gamma_{\ell m,I} +  \Gamma_{\ell m,S} +  \Gamma_{\ell m,T} \,.
\end{eqnarray} 

\subsection{Initial condition term and $q-$dependent anisotropies} 

Let us first evaluate the initial condition term

\begin{equation} 
\Gamma_{\ell m,I} = \int \frac{d^3 k}{\left( 2 \pi \right)^3} \,  {\rm e}^{i \vec{k} \cdot \vec{x}_0} \,  \Gamma \left( \eta_{\rm in} ,\, \vec{k} ,\, q  \right) \, 
\int d^2 n \,  Y_{\ell m}^* \left( {\hat n} \right) \,  {\rm e}^{-i k   \left( \eta_0 - \eta_{\rm in} \right) 
{\hat k} \cdot {\hat n} } \,.
\label{Gamma-lm-I}
\end{equation} 

Following the standard treatment for CMB anisotropies \cite{Dodelson:2003ft}, we make use of the identity 
\begin{eqnarray} 
{\rm e}^{-i {\bf k} \cdot {\bf y}} = \sum_\ell \, \left( - i \right)^\ell \left( 2 \ell + 1 \right) j_\ell \left( k y \right) P_\ell \left( {\hat k} \cdot {\hat y} \right) 
= 4 \pi \sum_\ell \sum_{m=-\ell}^\ell  \, \left( -i \right)^\ell \, j_\ell \left( k y \right) Y_{\ell m} ( {\hat k})\;  Y_{\ell m}^* \left( {\hat y} \right)\,, \nonumber\\ 
\end{eqnarray} 
(where $j_\ell$ and $P_\ell$ are, respectively, spherical Bessel functions and Legendre polynomial) so to obtain 
\begin{equation}
\Gamma_{\ell m,I} = 4 \pi \left( - i \right)^l \,  
\int \frac{d^3 k}{\left( 2 \pi \right)^3} \,  {\rm e}^{i \vec{k} \cdot \vec{x}_0} \, 
\Gamma \left( \eta_{\rm in} ,\, \vec{k} ,\, q  \right) \, Y_{\ell m}^* \left( {\hat k} \right) \, j_\ell \left( k \left( \eta_0 - \eta_{\rm in} \right) \right) \,.
\end{equation} 
Here $ \vec{x}_0$ denotes our location (that can be set to the origin), $\eta_0$ denotes the present time,  and $\eta_{in}$ the initial time. Once again we stress the peculiar property of the initial condition, namely its dependence on the frequency $q$. In  Section \ref{sec: GWcorrelator} we discuss how this imprints the SGWB angular spectrum.

\subsection{Scalar sourced term} 

A second source of anisotropy is due to the GW propagation in the large-scale scalar perturbations of the universe (the wavenumber of these perturbations $k$ is many order of magnitudes smaller than the GW frequency $q$, and the GW acts as a probe of this large-scale background).  As long as the scalar perturbation is in the linear regime (which is the case for the large-scale modes that leave an impact on the large-scale anisotropies of our interest), we can express it  \cite{Dodelson:2003ft} as a transfer function (a deterministic function that encodes the time-dependence of the perturbations) times a stochastic variable $\zeta$. This assumes 
the absence of isocurvature modes, and, in particular, of  anisotropic stresses, as for example those due to the relic neutrinos. This also assumes that the statistical properties of $\zeta$ have been set well before the propagation stage that we are considering (for instance during inflation, or during some early phase transition). Therefore, the scalar perturbations are 
\begin{equation} 
\Phi \left( \eta ,\, \vec{k} \right) = T_{\Phi} \left( \eta ,\, k \right) \zeta \left( \vec{k} \right) \;,\quad\quad \;\;\; 
\Psi \left( \eta ,\, \vec{k} \right) = T_{\Psi} \left( \eta ,\, k \right) \zeta \left( \vec{k} \right) \;. 
\end{equation} 
Under the above assumptions, $T_\Phi ( \eta , k ) =  T_\Psi ( \eta , k )$. However, we keep these two terms as distinct in our intermediate computations, so that the present analysis can be most easily generalized, if one wishes to introduce more general  sources. 

With this in mind, the scalar sourced term becomes
\begin{eqnarray}
\Gamma_S \left( \eta_0 ,\, \vec{k} ,\,  {\hat n} \right) &=&  \int_{\eta_{\rm in}}^{\eta_0} d \eta' \, {\rm e}^{i k \mu \left( \eta' - \eta_0 \right)} \left[ T_\Phi \left( \eta' ,\, k \right) \delta \left( \eta' - \eta_{\rm in} \right) + 
\frac{\partial \left[ T_\Psi \left( \eta' ,\, k \right) +  T_\Phi \left( \eta' ,\, k \right) \right] }{\partial \eta'} \right] \zeta \left( \vec{k} \right) \nonumber\\ 
&\equiv&   \int_{\eta_{\rm in}}^{\eta_0} d \eta' \, {\rm e}^{-i k \mu \left( \eta_0 - \eta' \right)} 
{\cal T}_S \left( \eta' ,\, k \right) \,  \zeta \left( \vec{k} \right) \,,
\label{scalarsolution}
\end{eqnarray} 
and we note that we are assuming a single adiabatic mode (i.e. $\zeta ( \vec{k} ) $ is the operator 
associated with the conserved curvature perturbation at 
 super-horizon scales). Proceeding as above, 
\begin{eqnarray} 
\Gamma_{\ell m,S} &=& 4 \pi \left( - i \right)^l \,  
\int \frac{d^3 k}{\left( 2 \pi \right)^3} \,  {\rm e}^{i \vec{k} \cdot \vec{x}_0} \, 
\zeta \left( \vec{k} \right) \, Y_{\ell m}^* \left( {\hat k} \right) \Bigg\{ 
T_\Phi \left( \eta_{\rm in} ,\, k \right) \, j_\ell \left( k \left( \eta_0 - \eta_{\rm in} \right) \right) \nonumber\\ 
&& \quad\quad\quad\quad  \quad\quad\quad\quad  + \int_{\eta_{\rm in}}^{\eta_0} d \eta' \, 
\frac{\partial \left[ T_\Psi \left( \eta' ,\, k \right) +  T_\Phi \left( \eta' ,\, k \right) \right] }{\partial \eta'} \, 
j_\ell \left( k \left( \eta_0 - \eta' \right) \right) \Bigg\}\,. \nonumber\\ 
\label{scalar}
\end{eqnarray} 
 As we can see, also the SGWB, feels, similarly to the CMB, a Sachs-Wolfe and integrated Sachs-Wolfe effect, which are represented by the first and the second term in \eqref{scalar}, respectively.

\subsection{Tensor sourced term}

Finally, the third contribution $\Gamma_{\ell m,T}$ is due to the GW propagation in the large-scale tensor modes 
\begin{eqnarray} 
\Gamma_{\ell m,T} &=&- \int d^2 n \,  Y_{\ell m}^* \left( {\hat n} \right) \int \frac{d^3 k}{\left( 2 \pi \right)^3} \, {\rm e}^{i \vec{k} \cdot \vec{x}_0} \, 
\frac{n^i \, n^j}{2} \,   \int_{\eta_{\rm in}}^\eta d \eta' \, {\rm e}^{i k \mu \left( \eta' - \eta_0 \right)} \, \frac{\partial \chi_{ij} \left( \eta' ,\, \vec{k} \right)}{\partial \eta'}\,.
\label{Gamma-lmT-1}
\end{eqnarray} 
To evaluate such term we decompose the tensor modes in right and left-handed  (respectively $\lambda=\pm2$) circular polarizations (see e.g. \cite{Bartolo:2018qqn}), 
\begin{equation}
\chi_{ij} \equiv \sum_{\lambda=\pm2} e_{ij,\lambda} \left( {\hat k} \right) \, \chi \left( \eta , k \right) \, \xi_\lambda \left( k^i \right) \,.   
\label{Gamma-lmT-2}
\end{equation} 
The three factors involved in each term are, respectively, the tensor circular polarization operator, the tensor mode function (equal for the two polarizations), and the stochastic variable for that tensor polarization (that is the analog of $\zeta$ we discussed in the scalar case). 

Inserting this decomposition in Eq. (\ref{Gamma-lmT-1}), a lengthy algebra, that we report in Appendix \ref{app:tensor-sourced}, leads to 
\begin{eqnarray} 
&& \!\!\!\!\!\!\!\! 
\Gamma_{\ell m,T} =   \pi  \left( - i \right)^\ell 
 \,  \sqrt{ \frac{\left(\ell +2 \right)!}{\left( \ell - 2 \right)!}} 
\int \frac{d^3 k}{\left( 2 \pi \right)^3} \, {\rm e}^{i \vec{k} \cdot \vec{x}_0} 
\sum_{\lambda=\pm2}  \,  _{-\lambda}Y_{\ell m}^* \left( \Omega_k \right) \, \xi_\lambda \left( \vec{k} \right) 
\, \int_{\eta_{\rm in}}^{\eta_0} d \eta \, \chi' \left( \eta ,\, k \right) \, 
 \frac{j_\ell \left( k \left( \eta_0 - \eta \right) \right)}{k^2 \left( \eta_0 - \eta \right)^2}  \,, \nonumber\\  
\label{Gamma-lmT-3}
 \end{eqnarray} 
which is formally analogous to Eq. \eqref{scalar}, with the product ${\hat \zeta} \, Y_{\ell m}^*$ replaced by the combination $\sum_{\lambda= \pm2}  {\hat \xi}_\lambda ( \vec{k} ) \, _{- \lambda}Y_{\ell m}^* ( \Omega_k ) $, involving the spin-2 spherical harmonics, and with the scalar transfer function replaced by the tensor one.

%%%%%%%%%%%
\subsection{Summary of the three contributions}
\label{subsec:GW-Gamma}
%%%%%%%%%%%

The results derived in the three previous subsections can be written in the (slightly) more compact form 
\begin{eqnarray}
\Gamma_{\ell m,I} \left( q \right)  &=& 4 \pi \left( - i \right)^l \,  
\int \frac{d^3 k}{\left( 2 \pi \right)^3} \,  {\rm e}^{i \vec{k} \cdot \vec{x}_0} \, 
\Gamma \left( \eta_{\rm in} ,\, \vec{k} ,\, q  \right) 
 Y_{\ell m}^* \left( {\hat k} \right) \, j_\ell \left( k \left( \eta_0 - \eta_{\rm in} \right) \right) \,,\nonumber
\\
\Gamma_{\ell m,S} &=&  4 \pi \left( - i \right)^l \,  
\int \frac{d^3 k}{\left( 2 \pi \right)^3} \,  {\rm e}^{i \vec{k} \cdot \vec{x}_0} \, 
\zeta \left( \vec{k} \right) Y_{\ell m}^* \left( {\hat k} \right) \, {\cal T}_\ell^S \left(  k ,\, \eta_0 ,\, \eta_{\rm in} \right)  \,,\nonumber
\\
\Gamma_{\ell m,T}  &=&  4 \pi \left( - i \right)^l \,  
\int \frac{d^3 k}{\left( 2 \pi \right)^3} \,  {\rm e}^{i \vec{k} \cdot \vec{x}_0} 
\sum_{\lambda=\pm2}  \,  _{-\lambda}Y_{\ell m}^* \left( \Omega_k \right) \, \xi_\lambda \left( \vec{k} \right) 
{\cal T}_\ell^T \left(  k ,\, \eta_0 ,\, \eta_{\rm in} \right)\,, 
\label{eq:solutionlm}
\end{eqnarray}
where  we have introduced the  linear transfer function ${\cal T}_\ell^{X(z)}$, with $X= S, T$ which represents the time evolution of the graviton fluctuations originated from the primordial perturbation
\begin{eqnarray}
{\cal T}_\ell^S \left( k ,\, \eta_0 ,\, \eta_{\rm in} \right) &\equiv& T_\Phi \left( \eta_{\rm in} ,\, k \right) \, j_\ell \left( k \left( \eta_0 - \eta_{\rm in} \right) \right)
 + \int_{\eta_{\rm in}}^{\eta_0} d \eta' \, 
\frac{\partial \left[ T_\Psi \left( \eta ,\, k \right) +  T_\Phi \left( \eta ,\, k \right) \right] }{\partial \eta} \, 
j_\ell \left( k \left( \eta - \eta_{\rm in} \right) \right) \,,\nonumber\\ 
{\cal T}_\ell^T \left( k ,\, \eta_0 ,\, \eta_{\rm in} \right) &\equiv&  \sqrt{ \frac{\left(\ell +2 \right)!}{\left( \ell - 2 \right)!}} \, \frac{1}{4} \int_{\eta_{\rm in}}^{\eta_0} d \eta \, \frac{ \partial \chi \left( \eta ,\, k \right)}{\partial \eta} \,  \frac{j_\ell \left( k \left( \eta_0 - \eta \right) \right)}{k^2 \left( \eta_0 - \eta \right)^2}\,. \,
\end{eqnarray}

%%%%%%%%%%%
\section{Correlators of GW anisotropies and SGWB non-Gaussianity}
\label{sec: GWcorrelator}
%%%%%%%%%%%

We now compute the 2-point $\langle \Gamma_{\ell m}   \Gamma_{\ell' m'}^*  \rangle$ and the 3-point $\langle \Gamma_{\ell_1 m_1}   \Gamma_{\ell_2 m_2} \Gamma_{\ell_3 m_3}  \rangle$ angular correlators of the solutions (\ref{eq:solutionlm}). The statistical operators entering in these solutions are the four momentum-dependent quantities $\Gamma ( \eta_{\rm in} ,\, \vec{k} ,\, q  ) ,\; \zeta ( \vec{k} ) ,\;  \xi_R ( \vec{k} ),\, {\rm and} \;  \xi_L ( \vec{k} )$, while the other terms encode deterministic effects such has the time evolution of the large-scale modes (in the linearized theory of the cosmological perturbations) and the projection of the GW anisotropies in the harmonic space. In this study, we assume that the stochastic variables are nearly Gaussian, with the 2-point  functions 
\begin{eqnarray}
\left\langle   \Gamma \left( \eta_{\rm in} ,\, \vec{k} ,\, q  \right)  \Gamma^* \left( \eta_{\rm in} ,\, \vec{k}' ,\, q  \right) \right\rangle & = & \frac{2 \pi^2}{k^3} \, P_{I} \left( q ,\, k \right) \, \left( 2 \pi \right)^3 \delta \left( \vec{k} - \vec{k}' \right) \,, \nonumber\\ 
\left\langle \zeta \left( \vec{k} \right) \zeta^* \left( \vec{k}' \right) \right\rangle &=& \frac{2 \pi^2}{k^3} \, P_\zeta \left( k \right) \, \left( 2 \pi \right)^3 \delta \left( \vec{k} - \vec{k}' \right) \;, \nonumber\\ 
\left\langle \xi_\lambda \left( \vec{k} \right) \xi_{\lambda'}^* \left( \vec{k}' \right) \right\rangle &=&  \frac{2 \pi^2}{k^3} \, P_\lambda \left( k \right) \, \delta_{\lambda,\lambda'}\, \left( 2 \pi \right)^3 \delta \left( \vec{k} - \vec{k}' \right)\,, 
\label{source-spectrum} 
\end{eqnarray} 
and a subdominant 3-point component 
\begin{eqnarray}
\left\langle   \Gamma \left( \eta_{\rm in} ,\, \vec{k} ,\, q  \right)  \Gamma^* \left( \eta_{\rm in} ,\, \vec{k}' ,\, q  \right) 
\Gamma^* \left( \eta_{\rm in} ,\, \vec{k}'' ,\, q  \right) \right\rangle & = & 
B_I \left( q ,\, k ,\, k' ,\, k''  \right) \, \left( 2 \pi \right)^3 \delta \left( \vec{k} + \vec{k}' + \vec{k}'' \right) \nonumber\\ 
\left\langle \zeta \left( \vec{k} \right) \zeta \left( \vec{k}' \right)  \zeta \left( \vec{k}'' \right) 
 \right\rangle &=& B_\zeta \left( k ,\, k' ,\, k'' \right) \, \left( 2 \pi \right)^3  \delta \left( \vec{k} + \vec{k}' + \vec{k}'' \right) 
\nonumber\\ 
\left\langle \xi_\lambda \left( \vec{k} \right) \xi_{\lambda'} \left( \vec{k}' \right)  \xi_{\lambda''} \left( \vec{k}'' \right) 
 \right\rangle &=&  B_\lambda \left( \vec{k} ,\, \vec{k}' ,\, \vec{k}'' \right) \, \delta_{\lambda,\lambda'}  \,  \delta_{\lambda,\lambda''}  
 \,  \left( 2 \pi \right)^3 \delta \left( \vec{k} + \vec{k}' + \vec{k}'' \right) \;. \nonumber\\ 
\label{source-bispectrum} 
\end{eqnarray} 
The assumption of nearly Gaussian modes is experimentally verified for the large-scale perturbations of $\zeta$ and of $\xi_\lambda$, as obtained from the CMB data \cite{Akrami:2019izv}. We assume that this is the case also for the initial condition term. 

The expressions (\ref{source-spectrum}) and (\ref{source-bispectrum}) can be readily used to compute the angular correlators of the solutions in (\ref{eq:solutionlm}). Moreover, for simplicity of exposition, we have here assumed that the various terms are not cross-correlated. These results in separate sets of correlators for the three terms in  (\ref{eq:solutionlm}). This assumption can be easily relaxed, and in fact, we did so in \cite{Bartolo:2019zvb} where we studied the anisotropic distribution of the GW originated in models with primordial black holes, as we review in Section \ref{sec: spectrald}. 

The computations performed so far assume statistical isotropy (recall the discussion at the end of Section \ref{sec: BoltzmannGW}). Correspondingly, when we combine  (\ref{source-spectrum}) and (\ref{source-bispectrum}) with  (\ref{eq:solutionlm}) we obtain angular correlators with well specific dependence on the multipole indices. Specifically, the two point correlators have the dependence 
\begin{eqnarray} 
\left\langle \Gamma_{\ell m}   \Gamma_{\ell' m'}^*  \right\rangle \equiv \delta_{\ell \ell'} \,\, \delta_{mm'} \, {\widetilde C}_\ell \;, \quad\quad\;\;\;\;\; \left\langle \Gamma_{\ell_1 m_1}   \Gamma_{\ell_2 m_2}    \Gamma_{\ell_3 m_3} \right\rangle \equiv 
\left( \begin{array}{ccc} 
\ell_1 & \ell_2 & \ell_3 \\ 
m_1 & m+2 & m_3   
\end{array} \right) \, {\widetilde b}_{\ell \ell' \ell''} \;, 
\label{Cell-Bell-def}
\end{eqnarray} 
while, under the above assumption, the angular power spectrum and the reduced bispectrum consists of the three separate contributions 
\begin{equation}
{\widetilde C}_\ell =   {\widetilde C}_{\ell,I} \left( q \right) + {\widetilde C}_{\ell,S} + {\widetilde C}_{\ell,T} \;, \quad\quad \;\;\;\;\;
{\widetilde b}_{\ell_1 \ell_2 \ell_3}  =   {\widetilde b}_{\ell_1 \ell_2 \ell_3,I}  \left( q \right) +  {\widetilde b}_{\ell_1 \ell_2 \ell_3,S}  +   {\widetilde b}_{\ell_1 \ell_2 \ell_3,T}  \,. 
\label{Cl-Bl}
\end{equation}   
We recall that the form of the bispectrum factorizes the Wigner-3j symbols \cite{Komatsu:2002db}, which are nonvanishing only provided that $\sum_i m_i=0$ and that the three $\ell_i$ satisfy the triangular inequalities. 

In the following we provide the explicit expression for the various contributions to the power spectrum and the reduced bispectrum.

\subsection{Angular power spectrum of GW energy density} 
\label{angularspec}

We start with the computation of the two-point function of the initial condition term. From the first of (\ref{eq:solutionlm}) we can write 
\begin{eqnarray}
\left\langle \Gamma_{\ell m,I} \left( q \right)  \Gamma_{\ell' m',I}^* \left( q \right) \right\rangle &=& 
\left( 4 \pi \right)^2 \left( - i \right)^{\ell - \ell'} 
\int \frac{d^3 k}{\left( 2 \pi \right)^3}  {\rm e}^{i \vec{k} \cdot \vec{x}_0} 
\int \frac{d^3 k'}{\left( 2 \pi \right)^3} {\rm e}^{-i \vec{k}' \cdot \vec{x}_0}  
\left\langle   \Gamma \left( \eta_{\rm in} ,\, \vec{k} ,\, q  \right)  \Gamma^* \left( \eta_{\rm in} ,\, \vec{k}' ,\, q  \right) \right\rangle\nonumber\\ 
& & \times ~ Y_{\ell m}^* \left( {\hat k} \right) Y_{\ell' m'} \left( {\hat k}' \right)
j_\ell \left(k \left( \eta_0 - \eta_{\rm in} \right) \right) j_{\ell'} \left(k' \left( \eta_0 - \eta_{\rm in} \right) \right) 
\,. 
\end{eqnarray} 
The correlator of the initial condition term is then given by the first of (\ref{source-spectrum}). Using this, and the orthonormality  condition of the spherical harmonics, $\int d ^2 {\hat n} \, _sY_{\ell m} \,_sY_{\ell' m'}^* = \delta_{\ell \ell'} \, \delta_{m m'} $, leads to 
\begin{eqnarray}
\left\langle \Gamma_{\ell m,I} \left( q \right)  \Gamma_{\ell' m',I}^* \left( q \right) \right\rangle &=& 
\delta_{\ell \ell'} \delta_{m m'}  \, 4 \pi \, \int \frac{d k}{k} \, \left[  j_\ell \left(k \left( \eta_0 - \eta_{\rm in} \right) \right) \right]^2  \,  
P_{I} \left( q ,\, k \right)  \,, 
\end{eqnarray}
which indeed is of the form dictated by statistical isotropy. The other two terms are obtained analogously. Altogether, we find 
\begin{eqnarray}
{\widetilde C}_{\ell,I} \left( q \right)   &=&  4 \pi \, \int \frac{d k}{k} \,  \left[  j_\ell \left(k \left( \eta_0 - \eta_{\rm in} \right) \right) \right]^2 
\,  P_{I} \left( q ,\, k \right) \,, \nonumber\\ 
{\widetilde C}_{\ell,S}  &=&  4 \pi  \int \frac{dk}{k} \,  {\cal T}_\ell^{\left( S \right) \,2} \left( k ,\, \eta_0 ,\, \eta_{\rm in} \right) 
\, P_{\zeta} \left( k \right)  \;, \nonumber\\ 
{\widetilde C}_{\ell,T}  &=&  4 \pi \int \frac{d k}{k} \, {\cal T}_\ell^{\left( T \right) \,2}   \left( k ,\, \eta_0 ,\, \eta_{\rm in} \right) \sum_{\lambda=\pm2}  P_\lambda \left( k \right)   \;. 
\label{Cell-res}
\end{eqnarray}
We know from the CMB that the large-scale tensor modes have a power smaller than the scalar ones. At large scale, the scalar contribution is dominated by the term proportional to the initial value of $\Phi$ in ${\cal T}_\ell^{(0)}$, which is the analog  of the SW contribution for the CMB. The large-scale modes that we are considering re-entered the horizon during matter domination. For these modes, ignoring the late time dark energy domination, $T_\Phi = T_\Psi = 3/5$ \cite{Dodelson:2003ft}. So, for scale invariant power spectra, 
\begin{equation}
{\widetilde C}_\ell \simeq {\widetilde C}_{\ell,I} \left( q \right) +  {\widetilde C}_{\ell,S} \simeq \frac{2 \pi}{\ell \left( \ell + 1 \right)} \left[ P_{I} \left( q \right) + \left( \frac{3}{5} \right)^2 \, P_{\zeta} \;\right] \;. 
\label{Cell}
\end{equation} 
The second term can be  compared to the SW contribution to the CMB anisotropies. In that case, the final temperature anisotropy is $1/3$ times the scalar perturbation at the last scattering surface, while $\Phi$ at that moment  decreased by a factor $9/10$ in the transition from radiation to matter domination  \cite{Dodelson:2003ft}. With this in mind, the second term in (\ref{Cell}) leads to $C_\ell^{\rm SW} = ( 3/10)^2 {\widetilde C}_{\ell,S}$, in agreement with the CMB literature.  On the other hand, if the two contributions are correlated, as it would be the case for adiabatic initial condition for $\Gamma_I$, then  both terms in  \eqref{Cell} contribute to the SW effect for the SGWB.

\subsection{Angular bispectrum  of GW energy density}
\label{sec: angbisp}

The characterization of the non-Gaussian properties of the SGWB is a potential tool to discriminate whether a SGWB has a primordial or astrophysical origin. The primoridal 3-point function of the GW field, $\left\langle h^3 \right\rangle$, is unobservable, due to the decoherence of the associated phase (because of the propagation, and the finite duration of the measurement \cite{Bartolo:2018evs,Bartolo:2018rku}), with, possibly, the exception of very specific shapes \cite{Dimastrogiovanni:2019bfl,Powell:2019kid}. It is more  convenient to consider the non-gaussianity associated to the GW energy density angular distribution, which is not affected by this problem \cite{Bartolo:2019zvb}. This gives rise to the bispectra in (\ref{Cl-Bl}), which we evaluate now. 

As we did for the power spectrum, also in this case we start from the initial condition term. Combining the first of (\ref{eq:solutionlm}) 
and the first of (\ref{source-bispectrum}) leads to 
\begin{eqnarray}
\left\langle  \prod_{i=1}^3 \Gamma_{\ell_i m_i,I} \left( q \right)  \right\rangle &=& 
\prod_{i=1}^3 \left[  4 \pi \left( - i \right)^{\ell_i} \,   \int \frac{d^3 k_i}{\left( 2 \pi \right)^3}  Y_{\ell_i m_i}^* \left( {\hat k}_i \right)  
j_{\ell_i} \left( k_i \left( \eta_0 - \eta_{\rm in} \right) \right) \right] \nonumber\\ 
&& \quad  \quad  \quad\quad 
\times  ~  B_I \left( q ,\, k_1 ,\, k_2 ,\, k_3 \right) \left( 2 \pi \right)^3 \delta^{(3)} \left( \vec{k}_1 + \vec{k}_2 + \vec{k}_3 \right)  \,. 
\end{eqnarray} 

We then use the representation of the Dirac $\delta-$function,  
\begin{eqnarray} 
\delta^{(3)} \left( \vec{k}_1 + \vec{k}_2 + \vec{k}_3 \right) &=& 
\int \frac{d^3 y}{\left( 2 \pi \right)^3} \, {\rm e}^{i \left( \vec{k}_1 + \vec{k}_2 + \vec{k}_3 \right) \cdot \vec{y}} \nonumber\\
&=& \int_0^\infty d y \, y^2 \int d \Omega_y \, 
\prod_{i=1}^3 \left[ 2 
\sum_{L_i M_i}  i^{L_i} \,  j_{L_i} \left( k_i \, y \right) \, Y_{L_iM_i}^* \left( \Omega_y \right) Y_{L_iM_i} \left( {\hat k}_i \right) 
\right] \,,  \nonumber\\ 
\label{delta} 
\end{eqnarray} 
and the orthonormality of the spherical harmonics, to arrive to 
\begin{eqnarray} 
&& \!\!\!\!\!\!\!\!  \!\!\!\!\!\!\!\! 
\left\langle \prod_{i=1}^3 \Gamma_{\ell_i m_i,I} \left( q \right)  \right\rangle =  {\cal G}_{\ell_1 \ell_2 \ell_3}^{m_1 m_2 m_3} \, \int_0^\infty d r \, r^2 \,  \prod_{i=1}^3 \left[ \frac{2}{\pi} \int d k_i \, k_i^2 j_{\ell_i} \left( k_i \left( \eta_0 - \eta_{\rm in} \right) \right) \, j_{\ell_i} \left( k_i \, r \right) \right] \,  B_I \left( q ,\, k ,\, k' ,\, k'' \right) \,, \nonumber\\ 
\end{eqnarray} 
where we have introduced the Gaunt integrals 
\begin{eqnarray} 
{\cal G}_{\ell_1 \ell_2 \ell_3}^{m_1m_2m_3} &\equiv&  \int d^2 {\hat n} \, Y_{\ell_1 m_1} \left( {\hat n} \right)  \, Y_{\ell_2 m_2} \left( {\hat n} \right)  \, Y_{\ell_3 m_3} \left( {\hat n} \right) \nonumber\\ 
&=& \sqrt{\frac{\left( 2 \ell_1 + 1 \right) \left( 2 \ell_2 + 1 \right) \left( 2 \ell_3 + 1 \right) }{4 \pi}} \, 
\left( \begin{array}{ccc} 
\ell_1 & \ell_2 & \ell_3 \\ 
0 & 0 & 0 
\end{array} \right) \, 
\left( \begin{array}{ccc} 
\ell_1 & \ell_2 & \ell_3 \\\ 
m_1 & m_2 & m_3 
\end{array} \right) \;. 
\label{gaunt} 
\end{eqnarray} 
We remark that also the bispectrum from the initial condition also generally as an ${\cal O} \left( 1 \right)$ dependence on the GW frequency. 

An analogous computation leads to the contribution from the scalar modes 
\begin{eqnarray} 
&& \!\!\!\!\!\!\!\!  \!\!\!\!\!\!\!\! 
\left\langle \prod_{i=1}^3 \Gamma_{\ell_i m_i,S} \right\rangle =  {\cal G}_{\ell_1 \ell_2 \ell_3}^{m_1 m_2 m_3} \, \int_0^\infty d r \, r^2 \,  \prod_{i=1}^3 \left[ \frac{2}{\pi} \int d k_i \, k_i^2 \, {\cal T}_{\ell_i}^S \left(  k_i ,\, \eta_0 ,\, \eta_{\rm in} \right)  \, j_{\ell_i} \left( k_i \, r \right) \right] \, B_\zeta \left( k ,\, k' ,\, k'' \right) \,.\nonumber\\ 
\end{eqnarray} 

For the tensor sourced contribution we have
\begin{eqnarray} 
\left\langle \prod_{i=1}^3 \Gamma_{\ell_i m_i,T} \right\rangle &=& \sum_{\lambda=\pm2} 
\prod_{i=1}^3 \left[ 4 \pi \left( - i \right)^{\ell_i} 
\int \frac{k_i^2 d k_i}{\left( 2 \pi \right)^3} \,  {\cal T}_{\ell,i}^T \left(  k_i ,\, \eta_0 ,\, \eta_{\rm in} \right)  \int d \Omega_{k_i} \, _{-\lambda} Y_{\ell_i m_i}^* \left( \Omega_{k_i} \right)  \right] \, 
\left\langle \prod_{i=1}^3 \xi_\lambda \left( \vec{k}_i \right) \right\rangle \,. \nonumber\\ 
\label{T2B-1}
\end{eqnarray} 

Following  \cite{Shiraishi:2010kd}, in Appendix \ref{app:tensor2B} we show that also this contribution can be cast in a similar form to the previous two terms: 
\begin{eqnarray} 
\left\langle \prod_{i=1}^3 \Gamma_{\ell_i m_i,T} \right\rangle &=&  {\cal G}_{\ell_1 \ell_2 \ell_3}^{m_1m_2m_3} 
\, \left[ \prod_{i=1}^3 4 \pi \left( - i \right)^{\ell_i} \int \frac{k_i^2 \, d k_i}{\left( 2 \pi \right)^3} {\cal T}_{\ell,i}^T \left(  k_i ,\, \eta_0 ,\, \eta_{\rm in} \right) \right]  \sum_{\lambda = \pm 2} {\tilde {\cal F}}_{\ell_1 \ell_2 \ell_3}^\lambda \left( k_1 , k_2 , k_3 \right)\,, \nonumber\\ 
\label{T2B-2}
\end{eqnarray} 
where 
\begin{eqnarray}
{\tilde {\cal F}}_{\ell_1 \ell_2 \ell_3}^\lambda \left( k_1 , k_2 , k_3 \right) & \equiv & \sqrt{4 \pi} 
\left( \begin{array}{ccc} 
\ell_1 & \ell_2 & \ell_3 \\ 
0 & 0 & 0 
\end{array} \right)^{-1} \, 
\sum_{m_1,m_2,m_3} 
\left( \begin{array}{ccc} 
\ell_1 & \ell_2 & \ell_3 \\ 
m_1 & m_2 & m_3  
\end{array} \right) 
\nonumber\\ 
&& \times ~ \left[ \prod_{i=1}^3 \int d \Omega_{k_i} \, \frac{  _{-\lambda}Y_{\ell_i m_i}^* \left( \Omega_{k_i} \right) }{\sqrt{2 \ell_i+1}} 
\right] \,  
\left\langle \xi_\lambda \left( \vec{k}_1 \right) \, \xi_\lambda \left( \vec{k}_2 \right) \, \xi_\lambda \left( \vec{k}_3 \right) 
\right\rangle\,. \nonumber\\ 
\label{cal-tilde-F}
\end{eqnarray} 

We remark once again that we have neglected for simplicity all the mixed scalar-tensor correlators.

 \subsection{Reduced Bispectrum and estimation}
 
The three contributions to the bispectrum found above have the correct form (\ref{Cell-Bell-def}) as dictated by statistical isotropy. 
For convenience, we collect here the explicit form of the reduced bispectra contributing to (\ref{Cl-Bl}) 
\begin{eqnarray} 
{\widetilde b}_{\ell_1 \ell_2 \ell_3,I} &=&  \int_0^\infty d r \, r^2 \,  \prod_{i=1}^3 \left[ \frac{2}{\pi} \int d k_i \, k_i^2 \,  j_{\ell_i} \left[ k_i \left( \eta_0 - \eta_{\rm in} \right) \right]  \;  j_{\ell_i} \left( k_i \, r \right) \right] \, B_{I} \left( q ,\, k_1 ,\, k_2 ,\, k_3 \right) \;, \nonumber\\ 
{\widetilde b}_{\ell_1 \ell_2 \ell_3,S} &=&  \int_0^\infty d r \, r^2 \,  \prod_{i=1}^3 \left[ \frac{2}{\pi} \int d k_i \, k_i^2 \, {\cal T}_{\ell_i}^S \left(  k_i ,\, \eta_0 ,\, \eta_{\rm in} \right)  \, j_{\ell_i} \left( k_i \, r \right) \right] \,  B_\zeta \left( k ,\, k' ,\, k'' \right)  \;, \nonumber\\ 
{\widetilde b}_{\ell_1 \ell_2 \ell_3,T}  &=& \frac{4}{\pi^2}  \sum_{\lambda = \pm 2} \sum_{m_i} 
 \left( \begin{array}{ccc} 
\ell_1 & \ell_2 & \ell_3 \\ 
0 & 0 & 0 
\end{array} \right)^{-2} \, {\cal G}_{\ell_1 \ell_2 \ell_3}^{m_1m_2m_3} 
\, \left[ \prod_{i=1}^3   \frac{ \left( - i \right)^{\ell_i}}{2 \ell_i+1} 
\int d^3 k_i {\cal T}_{\ell,i}^T (  k_i  ) 
\, _{-\lambda}Y_{\ell_i m_i}^* \left( \Omega_{k_i} \right)  \right] 
\nonumber\\ 
&& \quad\quad\quad \quad\quad\quad  \quad\quad\quad\quad  \quad\quad\quad\quad 
\times~  \delta \left( \vec{k} + \vec{k}' + \vec{k}'' \right) \, 
B_\lambda \left( \vec{k} ,\, \vec{k}' ,\, \vec{k}'' \right) \;. 
\label{Bell-res}
\end{eqnarray} 

To estimate the SGWB bispectrum, we consider only the scalar source contribution ${\widetilde B}_{\ell_1 \ell_2 \ell_3,S} $ 
and we assume the simplest small non-linear coupling {\it local} ansatz  for the curvature perturbation
\begin{equation}
\zeta \left( \vec{x} \right) = \zeta_g \left( \vec{x} \right) + \frac{3}{5} \, f_\NL \, \zeta_g^2 \left( \vec{x} \right) \,,
\end{equation} 
where $ \zeta_g \left( \vec{x} \right)$ denotes the linear Gaussian part of the perturbation. With the local ansatz, the bispectrum of the scalar perturbations assumes the form \cite{Gangui:1993tt,Komatsu:2001rj}
\begin{equation}
B_\zeta \left( k_1 ,\, k_2 ,\, k_3 \right) =  \frac{6}{5} f_\NL \left[
\frac{2 \pi^2}{k_1^3} P_\zeta \left( k_1 \right) 
\frac{2 \pi^2}{k_2^3} P_\zeta \left( k_2 \right) + 2 \; {\rm permutations} \right] \;. 
\end{equation} 
We insert this in the second of (\ref{Bell-res})  and we assume a matter transfer function $T_{\Phi} \left( \eta ,\, k \right) = T_{\Psi} \left( \eta ,\, k \right)=3/5\, g\left(\eta\right) $  with  the growth factor $g(\eta) = 1$  and a scale invariant spectrum for the primordial curvature fluctuations. We can then integrate over one of the internal momenta $k_i$, 
\begin{equation} 
\frac{2}{\pi} \, \int d k \, k^2 j_{\ell} \left( k \, \eta_0  \right) \, j_{\ell} \left( k \, r \right) \bigg | _{\ell \gg 1} = \frac{\delta \left( \eta_0 - r \right)}{\eta_0^2} \;. 
\label{integral-k}
\end{equation}
The relation (\ref{integral-k}) is exact if $k$ ranges up to infinity, which is not the case for the innermost momentum
(as the integral (\ref{integral-k}) is performed first, this will necessarily be the the momentum that we order to be the innermost one), due to the triangular inequalities associated to the bispectrum. 
The condition $\ell \gg 1$ ensures that the support of the integration occurs at sufficiently small $k$, so that the  relation (\ref{integral-k}) becomes exact at large $\ell$. The result then allows to immediately perform the integral over $r$. We then find that the reduced bispectrum from the scalar contribution, assuming that the SW is the dominant contribution,  is 
\begin{eqnarray} 
{\widetilde b}_{\ell_1 \ell_2 \ell_3,S} &=& \, 
\frac{162}{625} \, f_{\NL} \left(4\pi \, \int \frac{d k_1}{k_1} j_{\ell_1}^2 \left( k_1 \, \eta_0  \right) {\cal P}_{\zeta} \left( k_1 \right) \right) \, 
\left( 4 \pi \int \frac{d k_2}{k_2} j_{\ell_2}^2 \left( k_2 \, \eta_0  \right) {\cal P}_{\zeta} \left( k_2 \right)  \right)\nonumber\\
&&+ 2 \, {\rm permutations} \;. 
\end{eqnarray}

This result can also be written in terms of the two-point functions found in Eq. \eqref{Cell-res}: 
\begin{eqnarray}
{\widetilde b}_{\ell_1 \ell_2 \ell_3,S}&\simeq& 
 2 \, f_{\NL} \,  
\left[ \widetilde{C}_{\ell_1,S} \,   \widetilde{C}_{\ell_2,S} +   \widetilde{C}_{\ell_1,S} \,   \widetilde{C}_{\ell_3,S} +   \widetilde{C}_{\ell_2,S} \,   \widetilde{C}_{\ell_3,S} \right] \,, 
\label{reducedbis}
\end{eqnarray} 
which resembles the one for the CMB angular bispectrum in the Sachs-Wolfe regime \cite{Komatsu:2001rj}. So, the SGWB bispectrum is specified by the $f_\NL$ parameter and the angular spectrum.  Also in this estimate we neglected a possible correlation between the initial and scalar source contributions that should be taken into account when, for instance, $\Gamma_I$ is controlled by the adiabatic scalar perturbation (see \cite{Bartolo:2019zvb} for an example). 

\section{An example: the axion-inflation case} 
\label{sec: spectrald}

The goal of this section is to understand under which conditions the initial term $\Gamma_I \left( q \right)$ has a nontrivial $q-$dependence, that distinguishes it from the other contributions to the anisotropy. There are several mechanisms for the generation of a cosmological GW signal visible at interferometer scales (see \cite{Guzzetti:2016mkm,Bartolo:2016ami,Caprini:2018mtu} for recent review). In this section we focus on a  specific mechanism: we consider the case where an axion inflaton $\phi$ sources gauge fields, which in turn generates  a large GW background. In particular we consider the specific evolution shown in Figure 4 of \cite{Garcia-Bellido:2016dkw}, where the inflaton potential is chosen so to lead to a peak in the GW signal at LISA frequencies, without overproducing scalar perturbations and primordial black holes. 
The amount of GWs sourced in this mechanism is controlled by the parameter $\xi \equiv (\dot{\phi}/2 f_\phi H)$, where $f_\phi$ is the decay constant of the axion inflaton.
The present fractional energy in GW, $\Omega_\GW \left( \eta_0 ,\, q \right)$, is related to the primordial GW power spectrum $P_\lambda \left( \eta_{\rm in} ,\,  q \right)$ by 
\begin{equation}
\Omega_\GW \left( \eta_0 ,\, q \right) = \frac{3}{128} \, \Omega_{\rm rad} \, \sum_\lambda P_\lambda \left( \eta_{\rm in} ,\,  q \right)
\left[ \frac{1}{2} \left( \frac{q_{\rm eq}}{q} \right)^2 + \frac{4}{9} \left( \sqrt{2} - 1 \right) \right] \,.
\label{Omega-P}
\end{equation} 
This relation, taken from \cite{Caprini:2018mtu}, interpolates between large and small scales. Since we are interested in the modes with $q \gg q_{\rm eq}$, that entered the horizon during radiation domination, we consider only the second term in the square bracket, and we find
\begin{equation}
\Omega_\GW \left( \eta_0 ,\, q \right) = {\rm constant} \times  \sum_\lambda P_\lambda \left( \eta_{\rm in} ,\,  q \right) \,,
\end{equation} 
and, as we will see, the constant term is not relevant for our computation. 

We are interested in the contribution from the initial condition $\Gamma_{\rm in}$. So we can set the long modes $\zeta ( \vec{k} ) = h_\lambda ( {\hat k} ) = 0$ in this discussion. We therefore assume that the value of the energy density that arrives to the location $\vec{x}$ from the direction ${\hat n}$ is controlled by the parameter 
\begin{equation}
\xi = {\bar \xi} + \delta \xi \left( \vec{x} + d \, {\hat n} \right) \,.
\end{equation}
In this relation, $\xi$ is the value that this parameter had during inflation at the location $ \vec{x} + d \, {\hat n} $, where $d$ is the distance covered by the gravitons between the initial and the present time (equal for all directions, since we are disregarding the effect of  the long scale modes $\zeta$; we note that these modes will contribute to the term $\Gamma_S$, that we are not discussing in this section). In writing this relation, we have assumed that the parameter $\xi$ is in turn controlled by a dynamical field (the rolling axion, in the example of  \cite{Garcia-Bellido:2016dkw}), which results in the background value ${\bar \xi}$, and in the perturbation  $\delta \xi$. 

We then generalize the relation (\ref{Omega-P}) to 
\begin{equation}
\omega_\GW \left( \eta_0 ,\, \vec{x} ,\, q ,\, {\hat n} \right) =   {\rm constant} \times \sum_\lambda P_\lambda \left( q ,\, \xi \left( \eta_0 ,\, \vec{x} ,\, {\hat n} \right) \right) \,,
\end{equation} 
which has the background value ${\bar \Omega}_\GW \left( \eta_0 ,\, q \right) = {\rm constant} \times   \sum_\lambda P_\lambda \left( q ,\, {\bar \xi} \right)$.  The constant factor drops in the ratio 
\begin{equation}
4 -  \frac{\partial \ln \, {\bar \Omega}_\GW \left( \eta_0 ,\, q \right)}{\partial \ln \, q}   = 4 -   \frac{\partial \ln \left[  \sum_\lambda P_\lambda \left( q ,\, {\bar \xi} \right) \right]}{\partial \ln \, q} \;, 
\end{equation} 
as well as in 
\begin{equation}
\delta_\GW \left( \eta_0 ,\, \vec{x} ,\, q ,\, {\hat n} \right) = \frac{\sum P_\lambda \left( q ,\, \xi \left( \eta_0 ,\, \vec{x} ,\, {\hat n} \right) \right) - \sum P_\lambda \left( q ,\, {\bar \xi} \right)}{\sum P_\lambda \left( q ,\, {\bar \xi} \right)} = \frac{\partial \ln \left[ \sum_\lambda  {\rm ln } P_\lambda \left( q ,\, {\bar \xi} \right) \right]}{\partial {\bar \xi}} \, \delta \xi  \left( \vec{x} + d \, {\hat n} \right) \;,  
\end{equation} 
where we have expanded the GW  primordial power spectrum to linear order in $\delta \xi$. In this wat, the relation (\ref{delta-Gamma}) can be recast in the form 
\begin{equation}
\Gamma_I  ( \eta_0 ,\, \vec{x}_0 ,\, q ,\, {\hat n} ) \equiv {\cal F} ( q ,\, {\bar \xi} ) \, \delta \xi  ( \vec{x}_0 + d \, {\hat n} ) \;, 
\label{def-F}
\end{equation}
with 
\begin{eqnarray} 
{\cal F} \left( q ,\, {\bar \xi} \right)  \equiv \frac{1}{4-n_T} \, \frac{\partial \sum_\lambda \left[ \ln P_\lambda \left( q ,\, {\bar \xi} \right) \right]}{\partial {\bar \xi}} \;\;,\quad \;\; n_T \equiv    \frac{\partial \ln \left[  \sum_\lambda P_\lambda \left( q ,\, {\bar \xi} \right) \right]}{\partial \ln \, q} \;\;, 
\label{calF}
\end{eqnarray} 
where we have also made use of the standard definition of the tensor spectral tilt $n_T$. 

The question of whether we have or have not spectral distortion depends on whether the quantity ${\cal F} ( q ,\, {\bar \xi} ) $ is or is not  $q-$dependent. This provides an immediate criterion for evaluating whether and how much the GW anisotropies depend on frequency  
(as, in principle, one could imagine a GW power spectrum for which the dependence on $q$ of  ${\cal F}$ vanishes, or is extremely suppressed). This conclusion only assumes that the primordial GW signal is function of some additional parameter $\xi$ which has small spatial inhomogeneities, and therefore it likely applies to several other mechanisms. 

\begin{figure}[t!]
\centering 
\includegraphics[width=0.55\textwidth]{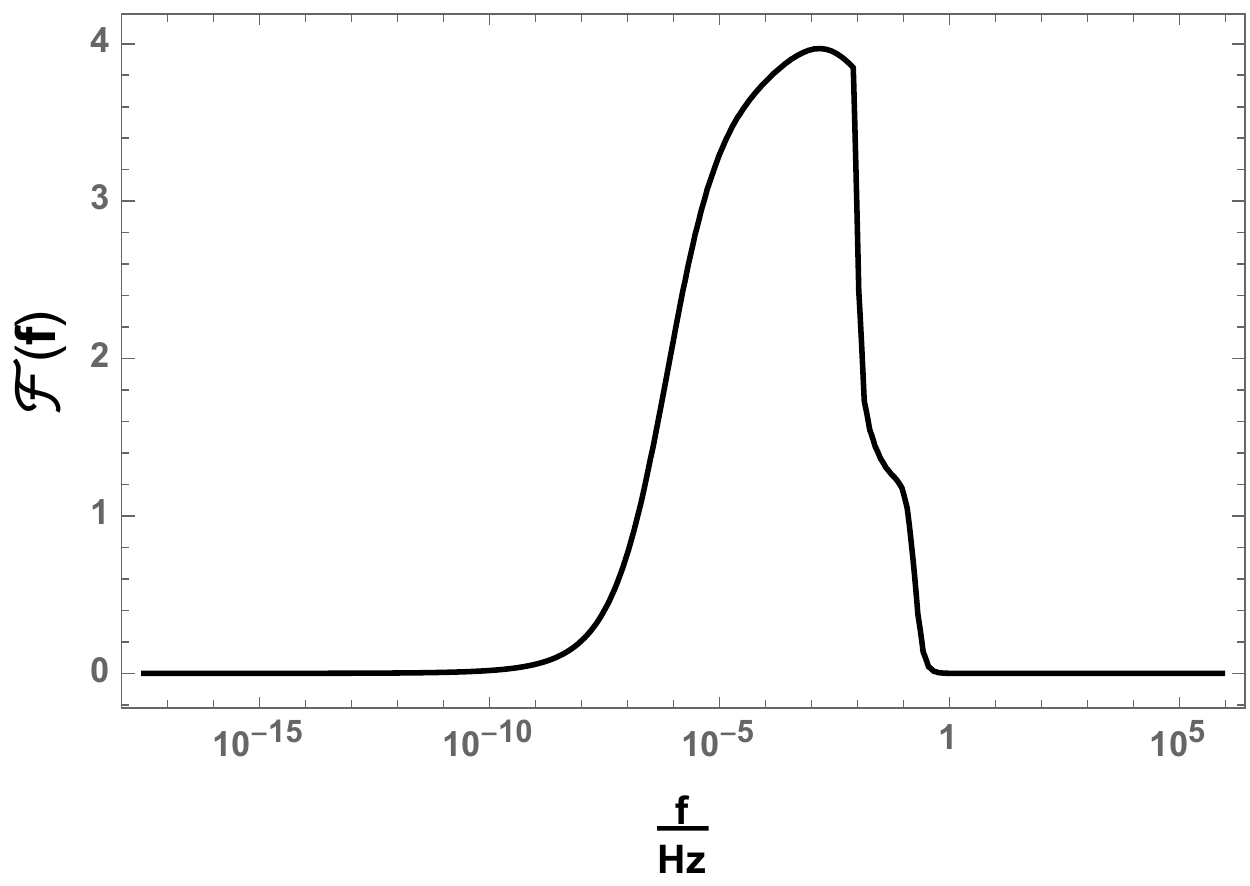}
\vskip -0.4cm
\caption{\it Quantity ${\cal F}$ as a function of the frequency $f = q/2 \pi$ of the GW signal for the model of axion inflation described in the text. }
\label{fig:F-sd}
\end{figure}

We show in Figure \ref{fig:F-sd} the evolution of the function ${\cal F}$ corresponding to the GW production shown in Figure 4 of  \cite{Garcia-Bellido:2016dkw}. We see that indeed this quantity presents a nontrivial scale dependence, and therefore the correlators of the anisotropies will be different at different frequencies. 

\section{Squeezed limit and consistency relations of the SGWB}
\label{sec-sqz}

Non-linear effects associated with the propagation of  interacting GWs  in a non-linear universe lead  to non-vanishing connected $n$-point functions even in absence of intrinsic, primordial non-Gaussianity. In particular, the squeezed
limit of  bispectra associated with GW observables should acquire a non-vanishing value, and satisfy consistency relations that resemble Maldacena's consistency relations \cite{Maldacena:2002vr}. This is analogous to what happens for CMB \cite{Bartolo:2011wb,Creminelli:2011sq,Lewis:2012tc}. 

\smallskip

In this Section we
 compute
the squeezed limit of the  bispectrum 
 for the graviton distribution function in the case of adiabatic fluctuations. 
 As in Section \ref{sec: BoltzmannGW}, we write in momentum space
 \begin{equation}
\omega_\GW \left(  \eta,\, k^i,\, q,\,n^j\right)
\,=\,\bar \omega_\GW \left( \eta,\, q\right)
\left[1+\delta_\GW  \left( \eta,\, k^i,\, q,\, n^i\right) \right]\,,
\end{equation}
where $\bar \omega_\GW ( \eta,\, q)$ is associated with the energy density of the isotropic  SGWB. This quantity depends on time $\eta$ and on the GW momentum $q$. 
Small anisotropies of the SGWB are controlled by 
the quantity $\delta_\GW$ given in Eq. \eqref{delta-Gamma}. We re-write it here, expressing it in terms of the function $\bar f(q)$:
%by
%\begin{equation}
%\delta_\GW  \left( \eta,\, k^i,\, q,\, n^i\right) \,\equiv\,\frac{\omega_\GW \left(  \eta,\, k^i,\, q,\,n^j\right)-\bar \omega_\GW \left( \eta,\, q\right)}{\bar \omega_\GW \left( \eta,\, q\right)}\,.
%\end{equation}
%This quantity, besides than on $\eta,\,q$, depends also on $k^i$ -- the Fourier momentum relative to a position $x^i$  -- and on $n^j$, %controlling the GW direction. It is convenient to parameterize
%  $\bar \omega_\GW$  in terms of a function $f(q)$ and use the relation \eqref{delta-Gamma} that we re-write here
\begin{eqnarray}
%\bar \omega_\GW(\eta,\,q)=\frac{q^4}{a^4(\eta)\,\rho_{\rm crit}}\,\bar f(q)\quad\quad {\rm and}\quad\quad
\delta_\GW(\eta,\,\vec k,\,q,\,\vec n)\,=\,-\frac{\partial \ln \bar f(q)}{ \partial\,\ln q}\,\Gamma_S\left( \eta,\,\vec k,\,q,\,\vec n\right)\;,\nonumber\\
\end{eqnarray}
where recall that $\Gamma_S$ controls the fluctuations in the distribution function (see the definitions in Section \ref{sec: BoltzmannGW}). In this Section we focus on the contribution due to scalar fluctuations.
 %
% In this Section we focus on the contribution due to scalar fluctuations, and from now on we write $\Gamma \,\equiv\,\Gamma_S$.\\ 
We assume there is no anisotropic stress, and that scalar perturbations in Newtonian gauge  satisfy the  adiabaticity condition:
% Then we can write
\begin{equation}\label{relSperts}
\Phi(\eta, \vec k)\,=\,\Psi(\eta,\vec k)\,=\,\frac35\,g(\eta)\,\zeta({\vec k})\,,
\end{equation}
 where 
 %$\zeta_{\vec k}$  is a time-independent operator defined at superhorizon scales, that  determines the statistical properties
%of the fluctuations. 
$g(\eta)$ is a function mapping the superhorizon seed (controlled by  $\zeta({\vec k})$) to the scalar fluctuations at small scales (see e.g. \cite{Lahav:1991wc,Carroll:1991mt}). It is generally time
dependent although it is equal to unity in pure matter domination.
Then the contribution  $\Gamma_{\rm S}$ reads (see eq \eqref{scalarsolution}) 
 \begin{eqnarray}
 \label{defGsc}
 \Gamma_{\rm S}(\eta,\vec k,\,\hat n)&=&
 \frac35\,
 \int_{\eta_{\rm in}}^{\eta}\,d \eta' \,e^{-i k \mu \left(\eta-\eta'\right)}\,
 \left[ \delta(\eta'-\eta_{\rm in}) \,g(\eta')+\frac12\,\partial_{\eta'}g(\eta')\right]
%{\cal T}_S (\eta',\,k)
\,\zeta({\vec k})\,,\nonumber
\\
&\equiv& {\it T}_S(\eta,\,k,\,\mu)\,\zeta \left( {\vec k} \right) \,,
 \end{eqnarray}
 where $\mu\,=\,\hat n\cdot\hat k\,$
and ${\it T}_S$ is the definition for the scalar transfer function we adopt here. In matter domination this becomes
\begin{equation}
{\it T}_S=\frac35\, e^{-i k \mu \left(\eta-\eta_{\rm in} \right)}\,.
\end{equation}
Notice that $ \Gamma_{\rm S}$ does not depend on $q$.
Assembling the definitions above, we can then write
\begin{eqnarray}
\delta_\GW(\eta,\,\vec k,\,q,\,\vec n)&=&-\frac{\partial \ln \bar f(q)}{ \partial\,\ln q}\, {\it T}_S(\eta,\,k,\,\mu)\,\zeta({\vec k})\,.
\end{eqnarray}
Indicating with $P_{\Gamma}$ the power spectrum, we can write the 2-point correlators in momentum space as
\begin{eqnarray} \label{Ga2pt}
{\left\langle \Gamma(\eta, \,\vec k_1,q\,,\,\hat n) \,\Gamma(\eta, \,\vec k_2,q\,,\,\hat n) \right\rangle}'&=&
\frac{2 \pi^2}{k_1^3}\,P_{\Gamma} (\eta, k_1,\,q,\,\hat n)\,\nonumber
\\
&=&
\frac{2 \pi^2}{k_1^3}\,\left| {\it T}_S(\eta,\,k_1,\,\mu_1) \right|^2\,P_{\zeta} ( k_1)\,,
\end{eqnarray}
where a prime $'$ corresponds to  correlators understanding the $(2 \pi)^3\,\delta(\sum \vec k_i)$ factor. Then:
\begin{eqnarray}\label{defGps}
P_{\Gamma} ( \eta,\,k,\,\mu)&=&
\frac{{\left\langle \Gamma_S ( \eta,\,k,\,\mu) \Gamma_S ( \eta,\,k',\,\mu) \right\rangle}'}{2 \pi^2/k^3}\,=\,
\left| {\it T}_S(\eta,\,k,\,\mu) \right|^2\,P_{\zeta} ( k)\,,
\\
P_{\delta_\GW} ( \eta,\,k,\,q,\,\mu)&=&\,\left| \frac{\partial \ln \bar f(q)}{ \partial\,\ln q}\,{\it T}_S(\eta,\,k,\,\mu) \right|^2\,P_{\zeta} ( k)\,.\label{defDps}
\end{eqnarray}
In matter domination, as we learned above, $\left| {\it T}_S\right|^2\,=\,9/25$, but in general  $\left| {\it T}_S\right|^2$  can depend on $\eta$, $k$, $\hat n$.\\

In what follows, we study how the two-point correlation functions of SGWB anisotropies, when evaluated at small scales $k$,  are modulated by 
the presence of a long-scale mode $\zeta_L\,\equiv\,\zeta({\vec k}_L)$, with $|\vec k_L|\,\ll\,|\vec k|$.  Such modulation induces a non-vanishing squeezed limit
for the three-point function of $\delta_\GW$. The anisotropies $\delta_\GW$ depend on
various quantities, $( \eta,\,k,\,q,\,\mu)$, which can be sensitive in a different way to the long mode. We use the systematic approach
pionereed by Weinberg \cite{Weinberg:2003sw} that unambiguously associates the effects of a long mode with an appropriate coordinate
transformation.  We shall closely follow the treatment of  \cite{Creminelli:2011sq}, which develops the arguments of   \cite{Weinberg:2003sw} 
for the case of CMB, applying it to the SGWB (for similar approaches see also~\cite{Bartolo:2011wb,Lewis:2012tc}).

%%%%%%%%%%%%%%%%%%%%%%%%%%%%%%%%%%
\subsection{Long wavelength modes as coordinate transformations}  \label{sec-coord-trans}
%%%%%%%%%%%%%%%%%%%%%%%%%%%%%%%%%%
   
We discuss how to identify the effects of a long mode with an appropriate coordinate transformation.  We limit our attention to effects due to scalar fluctuations. The metric including long-wavelength scalars in Poisson gauge is
\begin{equation}
d s^2\,=\,a^2(\eta) \left[ 
-\left(1+2 \,\Phi_ L\right)\,d \eta^2+\left(1-2 \Psi_L \right) \delta_{ij} \,d x^i d x^j 
\right]\,.
\end{equation}
 We assume that the long-scale mode depends on a momentum $\vec k_L$, with magnitude much smaller than that of the momentum of the short-scale modes introduced in eq. (\ref{line-short}), but with a certain direction, and we discuss how the quantities $( \eta,\,k,\,q,\,\mu)$,  transform under a coordinate redefinition adsorbing the long modes.  We start by noticing that the following coordinate transformation preserves the Poisson
 gauge structure ($\zeta_L$ indicates the long mode of curvature fluctuations at large scales, responsible for
 the modulation effects):
 \begin{eqnarray}
 \hat  \eta&=&\eta+\epsilon(\eta)\,\zeta_L \label{sparesc1}\,,
 \\
 \hat x^i&=& x^i \left(1-\lambda\,\zeta_L \right) \label{sparesc}\,,
 \end{eqnarray}
 with $\lambda$ constant. After  performing such gauge transformation,
 \begin{eqnarray}
 \hat \Phi_L&=& \Phi_L-\epsilon'\,\zeta_L-{\cal H}\,\epsilon\,\zeta_L\,,\nonumber
 \\
  \hat \Psi_L&=& \Psi_L-\lambda\,\zeta_L+{\cal H}\,\epsilon\,\zeta_L\,,
 \end{eqnarray}
  we can `gauge away' the long wavelength scalar modes making the gauge choice
 \begin{eqnarray}
 \Phi_L&=&\left(\epsilon'+{\cal H} \epsilon\right)\,\zeta_L\;,\nonumber
 \\
  \Psi_L&=&\left(\lambda-{\cal H} \epsilon\right)\,\zeta_L\;,
 \end{eqnarray}
 so that in the hat coordinates the metric is purely FRW with no long-wavelength perturbations. 
As explained in \cite{Weinberg:2003sw,Creminelli:2011sq}, in order  to be consistent  with the small $k$
limit of Einstein equations, we need
to impose the conditions  (in absence of anisotropic stress)
\begin{eqnarray}
\lambda&=&1\,, \nonumber
\\ \label{defeps}
\epsilon(\eta) &=&\frac{1}{a^2(\eta)} \,\int^\eta_{\eta_\star}\,d \eta'\, a^2(\eta')\, ,
\end{eqnarray}
where $\eta_*$ is some initial reference time. 
Eq \eqref{defeps} immediately leads to the
% Then, we have the 
 equality
\begin{equation}
\epsilon'\,=\,-2{\cal H}\,\epsilon+1\;.
\end{equation}
After performing the  coordinate redefinition \eqref{sparesc1},  \eqref{sparesc}, we can write 
a   metric containing short-wavelength scalar fluctuations in Poisson  gauge `on top' of long fluctuations:
\begin{equation}
d s^2\,=\,a^2(\hat \eta) \left[ 
-\left(1+2 \,\hat \Phi_ S\right)\,d \hat \eta^2+\left(1-2 \hat \Psi_S \right) \delta_{ij} \,d \hat x^i d \hat  x^j 
\right]\;.
\label{line-short}
\end{equation}
In fact, 
 such metric contains the long-scale modes within the definition  of the hat coordinates. We can then express the 
perturbations in 
 terms of the original coordinates $(\eta, \,x^i)$ using again relations  \eqref{sparesc1},  \eqref{sparesc}.
%without  the hats that
 %include the long modes. 
  Such operation  teaches us how the short wavelength modes are modulated
by the long wavelength ones:
\begin{eqnarray}
\hat \Phi_S &=&\Phi_S +\Phi_L+2 \Phi_S \Phi_L+\epsilon \,\zeta_L\,\frac{\partial \Phi_S}{\partial \eta}-\lambda \,\zeta_L\,
\,x^i \,\frac{\partial \Phi_S}{\partial x^i}\,,
\\
\hat \Psi_S &=&\Psi_S +\Psi_L-2 \Psi_S \Psi_L  +\epsilon \,\zeta_L\,\frac{\partial \Psi_S}{\partial \eta}-\lambda\,\zeta_L\,
\,x^i \,\frac{\partial \Psi_S}{\partial x^i}\,.
\end{eqnarray}
Importantly, the short modes acquire a second order correction due to long modes.  As we shall
discuss in what comes next, these non-linear,
higher-order corrections    modulate  the 2-point function
 for short modes, and lead to  a
 non-vanishing  squeezed limit for the 3-point function.

\smallskip
As a concrete example, that we shall use in what follows, we can consider the  
    case of constant proportionality  between pressure and energy density, $p=w \rho$. Being in this case
$a(\eta)\,\propto\,\eta^{2/(1+3 w)}$, 
$
{\cal H}\,=2/[\eta(1+3 w)]$ we get 
\begin{equation}
\epsilon(\eta)\,\zeta_L\,=\,\frac{1+3 w}{5+3 w}\,\eta\,\zeta_L\, , 
\end{equation}
and 
\begin{equation}
{\cal H}\,\epsilon\,=\,\frac{2}{5+3 w}\,,
\end{equation}
which, for matter domination, gives ${\cal H}\,\epsilon\,=\,2/5$. 

\smallskip
%\subsection{Transforming the  Fourier momentum $\vec k$}
We also   need
 to evaluate how the Fourier transform of a function $f(x^i)$ changes under a rescaling of spatial coordinates, as
 in eq \eqref{sparesc}. We find that if we apply a constant rescaling of spatial coordinates
 \begin{equation}
 f(x^i)\,\to\,f\left(x^i (1-\lambda\,\zeta_L) \right)
 \end{equation}
 to a function $f$, then  its  Fourier transform, given by
 \begin{eqnarray}
 f(x^i)&=& \int \frac{d^3 k}{(2 \pi)^3}\,e^{i \vec k \vec x} \tilde f(k^j), \nonumber
 \end{eqnarray}
 transforms as (at first order in a $\zeta_L$ expansion)
  \begin{eqnarray}
  f(x^i (1-\lambda\,\zeta_L))&=& \int \frac{d^3 k}{(2 \pi)^3}\,e^{i \vec k \vec x  (1-\lambda\,\zeta_L)} \tilde f(k^j)
   \,=\, 
 \int \frac{d^3 k}{(2 \pi)^3}\,e^{i \vec k \vec x } \left[
    (1+3\lambda\,\zeta_L) \tilde f(k^j  (1+\lambda\,\zeta_L)) \right]\;.\nonumber\\
\end{eqnarray}
This implies that up to first order in $\zeta_L$, under the coordinate transformation we are interested in, we have:
\begin{eqnarray}
 \tilde f (k^j) \to \left(1+3 \lambda\,\zeta_L \right)\,\tilde f \left( k^j (1+ \lambda\,\zeta_L ) \right) \hskip0.2cm\Rightarrow 
 \hskip0.2cm
  \tilde f (k^j) \to  \tilde f (k^j) +3 \lambda \,\zeta_L \tilde f (k^j) + \lambda\,\zeta_L\, k^m \,\frac{\partial \tilde f (k^j)}{\partial k^m}\;.\nonumber\\
 \end{eqnarray}

%\subsection{Transformation of the GW four-momentum}

As a last step, we now investigate how to transform the coordinates $(q,\,\hat n^i )$ that control the GW four-momentum.
  At first order, neglecting tensors, the GW four-momentum  components are given by 
  
    \begin{equation}
  P^0=\frac{q}{a^2(\eta)}\,e^{- \Phi},\quad\quad\quad\quad \quad
  P^i=\frac{q}{a^2(\eta)}\, n^i\,e^{\Psi}\,.
  \end{equation}
We wish to express the previous quantities in terms of hat coordinates, including the
effects of the long modes. In particular,  we are interested in determining the quantities $\hat q$ 
and $\hat n^i$ that are contained into the  GW four-momentum, when it is expressed in terms of hat coordinates.
%
 %in the 
%
%entering in the following definition for $\hat P^0$: 
%${\hat P}^0\,=\,\left({\hat q}/{a^2(\hat \eta)} \right)\,e^{- \hat \Phi}$, that contains the effects of the long modes. (We can define analogously for $\hat n^i$ from the vector $\hat P^i$.)
 We use
the fact that $P^\mu$ is a vector, transforming in the usual way under coordinate transformations (in
particular transformations \eqref{sparesc1}, \eqref{sparesc}). Using this fact, we find
%
%Since $P^\mu$ is a vector, it transforms
%in the usual way under a coordinate transformation
%\begin{eqnarray}
%P^\mu (x^\rho) \,=\,\frac{\partial  x^\mu}{\partial \hat x^\nu}\, P^\nu (\hat x^\rho)\,.
%\end{eqnarray} 
%The
%time and space components  of the hat quantity $\hat P^\mu$ transform as
\begin{eqnarray}
\frac{\hat q}{a^2 (\hat \eta)}%e^{-  \hat{\Phi}_S(\hat \eta, \hat x)}&=&\hat P^0\,=\,\left(1 +\epsilon' \,\zeta_L\,\right) P^0\,e^{-\Phi_L}
&=&\left(1 +\epsilon' \,\zeta_L\,\right)\,\frac{q}{a^2(\eta)}\,e^{-\Phi_L}\;,
\label{cond1} 
\\
\frac{\hat q}{a^2(\hat \eta)}\, \hat n^i\,%e^{ \hat{\Psi}_S(\hat \eta, \hat x)}
&=&
%\hat P^i\,=\,\left(1-\lambda\,\zeta_L\, \right) P^i\,=\,
\left(1-\lambda \,\zeta_L\,\right)\,\frac{q}{a^2(\eta)}\, n^i\,e^{\Psi_L}\;.
\label{cond2}
\end{eqnarray}
Condition \eqref{cond1} gives, at first order in the long-scale  modes, 
\begin{eqnarray}
\hat q &=&\frac{a^2(\hat \eta )}{a^2(\eta)}\, \left(1 +\epsilon' \,\zeta_L\,\right)\,\left(1-\Phi_L \right)\,q \nonumber
 \\
  &=&\left(1+2\,{\cal H}\,\epsilon \,\zeta_L\,+\epsilon' \,\zeta_L\,-\Phi_L\right)\,q \nonumber
\\
&=&\left[1+\left( 1-\frac{3}{5} g(\eta)\right)\,\zeta_L
\right]\,q\,.
 \end{eqnarray}
 On the other hand, condition \eqref{cond2} gives
 \begin{eqnarray}
 \hat n^i&=&\frac{a^2(\hat \eta )}{a^2(\eta)}\, \frac{q}{\hat q} \,\left(1-\lambda\,\zeta_L\right) \,\left( 1+\Psi_L\right) \,n^i\nonumber
\\
&=& \left( 1- \epsilon' \,\zeta_L\,-\lambda\,\zeta_L+2 \Phi_L\right) \,n^i\nonumber
\\
&=&\left[ 1\,-2\left( 1- {\cal H}\,\epsilon(\eta)  -\frac{3}{5} g(\eta)\right)\,\zeta_L  \right] n^i\, .
\end{eqnarray}
These are the results that we need.  
 It is convenient to write more compact expressions as
\begin{eqnarray}
\hat q &=&\left(1+{\beta}_q(\eta)\, \zeta_L\right) \,q\,, \nonumber
\\
\hat n^i&=&\left(1+{\beta}_n(\eta)\, \zeta_L\right) \,n^i\,,
\end{eqnarray}
with $\beta_{q,n}$ functions of time
\begin{eqnarray}
{\beta}_q(\eta)&=& 1-\frac{3}{5} g(\eta)\,, \nonumber
\\
{\beta}_n(\eta)&=&-2\left( 1- {\cal H}\,\epsilon(\eta)  -\frac{3}{5} g(\eta)\right)\,.
\end{eqnarray}
In matter domination  we find ${\beta}_q\,=\,2/5$ and ${\beta}_n\,=\,0$.
%\begin{eqnarray}
 %\hat q  &=&\left(1+\frac25 \zeta_L\right) \,q \hskip1cm {\text{(MD)}}
%\\
%\hat n^i &=&n^i \hskip2.9cm {\text{(MD)}}
%\end{eqnarray}

%%%%%%%%%%%%%%%%%%%%%%%%%%%
\subsection{Coordinate transformations and the GW distribution function}
%%%%%%%%%%%%%%%%%%%%%%%%%%%%%%

We now apply the previous results to the problem at hand. We start by re-writing the GW energy density
\begin{equation}
\omega_\GW \left(  \eta,\, k^i,\, q,\, n^i\right)
\,=\,\bar \omega_\GW \left( \eta,\, q\right)
\left[1+\delta_\GW  \left( \eta,\, k^i,\, q,\, n^i\right) \right]\,,
\end{equation}
where
\begin{eqnarray}
\bar \omega_\GW(\eta,\,q)&=&\frac{q^4}{a^4(\eta)\,\rho_{\rm crit}}\,\bar f(q) \;, 
\end{eqnarray}
and
\begin{eqnarray}
\delta_\GW(\eta,\,\vec k,\,q,\, n^i)&=&-\frac{\partial \ln \bar f(q)}{ \partial\,\ln q}\,\Gamma_S\left( \eta,\,\vec k,\,q,\, n^i\right) \label{defDelta}\,.
\end{eqnarray}
We now transform 
each contribution in the previous formulas under the  coordinate transformation discussed in Section \ref{sec-coord-trans}.
The background quantities $\bar \omega_\GW$ and $\bar f(q)$ transform as
\begin{eqnarray}\label{zerotr}
\bar \omega_\GW( \eta,\, q)
&\Rightarrow&
\bar \omega_\GW(\hat \eta,\,\hat q)\,=\,\bar \omega_\GW( \eta,\, q)
\left[1+4 \left( \beta_q-{\cal H}\,\epsilon\right)\,\zeta_L+ \beta_q\,\frac{\partial \ln \bar f(q)}{ \partial\,\ln q}\,
\zeta_L \right] \;, 
\end{eqnarray}
where
\begin{eqnarray}
\frac{\partial \ln \bar f(q)}{ \partial\,\ln q}&\Rightarrow&\frac{\partial \ln \bar f(\hat q)}{ \partial\,\ln \hat q}
\,=\,\frac{\partial \ln \bar f(q)}{ \partial\,\ln q}+\beta_q(\eta)\,\frac{\partial^2 \ln \bar f(q)}{ \partial\,(\ln q)^2}\,\zeta_L\,.
\end{eqnarray}
The quantity $\Gamma_S$ is mapped to
\begin{eqnarray}
 \Gamma_S\left( \hat \eta,\, \hat  k^i,\,\hat q,\,\hat  n^i\right)
\,=\,\left(1+3 \zeta_L \right)\,\Gamma_S\left(  \eta+\epsilon(\eta) \,\zeta_L\,,\,  \vec k \left(1+\zeta_L \right),\, \left(1 +\beta_q(\eta)\,\zeta_L \right)\,q,\,  \left(1 +\beta_n(\eta)\zeta_L\right)\, n^i\right)\,,\nonumber\\
\end{eqnarray}
that, 
expanded at linear order in $\zeta_L$, becomes
\begin{eqnarray}
\Gamma_S\left(  \eta,\,   k^i,\, q,\,  n^i\right)&\Rightarrow&
 \Gamma_S\left( \hat \eta,\, \hat  k^i,\,\hat q,\,\hat  n^i\right)\,=\,\left(1+3 \zeta_L \right)\, \Gamma_S\left(  \eta,\,   k^i,\, q,\,  n^i\right) \nonumber
\\
&&+\frac{\partial \Gamma_S}{\partial \,\eta}\,\epsilon(\eta)\,\zeta_L+k^i \,\frac{\partial \Gamma_S}{\partial \,k^i}\,\zeta_L+\beta_q(\eta)
\frac{\partial \Gamma_S}{\partial \ln q}\,\zeta_L+\beta_n(\eta)\,n^j \,\frac{\partial \Gamma_S}{\partial \,n^j}\,\zeta_L\,.
\nonumber\\
\end{eqnarray}
We now assemble
  the results obtained. The  SGWB energy density, including anisotropies,
  is modulated by the long mode 
$\zeta_L$
as
\begin{eqnarray}
 \omega_\GW \left( \hat \eta,\, \hat k^i,\, \hat q,\, \hat n^i\right)
 &=&
\bar \omega_{GW}
\left(  \eta,\, q \right)
\,\Bigg[1+
\delta_\GW +\left( 
4  \beta_q-4 \,{\cal H}\,\epsilon
 + \beta_q\,\frac{\partial \ln \bar f(q)}{ \partial\,\ln q}\,
\right)\zeta_L \nonumber
\\
&&+
\left(3  +  \beta_q(\eta)\,\frac{\partial \ln q}{ \partial \ln \bar f(q)} \frac{\partial^2 \ln \bar f(q)}{ \partial\,(\ln q)^2} \right)\,\zeta_L\,\delta_\GW + \nonumber
\\
&&+
 \left(\frac{\partial  \ln \Gamma}{\partial \,\eta}\,\epsilon(\eta)+k^i \,\frac{\partial  \ln \Gamma}{\partial \,k^i}\,
  +\beta_n(\eta) \,\frac{\partial \ln \Gamma}{\partial \,\ln \mu} \right)\,\zeta_L \,\delta_\GW \Bigg]\,.
\label{modgenga}
\end{eqnarray}
Eq. \eqref{modgenga}
  is the basic expression that we need: all quantities at the RHS are evaluated in terms of the original coordinates without the hat.
  Notice that {\it even in absence  of intrinsic small-scale anisotropies}, the GW energy density is modulated by the long mode: a dependence on $\zeta_L$ is indeed still present
  by setting $\delta_\GW\,=\,0$ in eq  \eqref{modgenga}. This is the effect studied by Alba and Maldacena \cite{Alba:2015cms}. For example, in pure matted domination, we have $\beta_q\,=\,{\cal H}\,\epsilon\,=\,2/5$. Setting $\delta_\GW\,=\,0$, eq  \eqref{modgenga}  simply becomes
\begin{eqnarray}
\hat \omega_\GW \left( \hat \eta,\, \hat k^i,\, \hat q,\, \hat n^i\right)
&=&
\bar \omega_{GW}
\left(  \eta,\, q \right)\,\left(1
+\frac25\,\frac{\partial \ln \bar f(q)}{ \partial\,\ln q}\,
 \zeta_L \right)\,.
\end{eqnarray}
In this case, 
the modulation of $ \omega_\GW$ is then controlled by the momentum dependence of the function $\bar f(q)$, associated with the isotropic distribution function of the SGWB energy density \cite{Alba:2015cms}.

\subsection{The squeezed limit of 3-point correlation functions}
We now apply the general result of   \eqref{modgenga} to study how correlation functions of small-scale GW anisotropies $\delta_\GW$ are influenced by the long mode. We start by studying how two-point correlation functions are modulated by $\zeta_L$; we continue investigating the squeezed limit of the three-point correlation functions.

\smallskip
%\subsection{Modulation of two-point correlation function}\label{sec_mod_2pt}
Using
eq \eqref{modgenga}, we  find the result\footnote{Each quantity is evaluated at the same value of $\eta$, $q$, $ n^i$, hence
we understand such dependence.  Here we indicate with $\hat \delta_\GW$ the quantity that receives the long-mode modulation.}
\begin{eqnarray}\label{tpmod1}
{\left\langle \hat \delta_\GW(\vec k_1)
\,\hat \delta_\GW(\vec k_2) \right\rangle}'&=&
\left(1+{\it {\cal M}}\,\,\zeta_L \right)
\,
{\left\langle  \delta_\GW(\vec k_1)
\, \delta_\GW(\vec k_2) \right\rangle}' \,,
\end{eqnarray}
 where the modulating factor ${\cal M}$ reads
\begin{eqnarray}
{\cal M}=&&6 +  2\beta_q(\eta)\,\frac{\partial \ln q}{ \partial \ln \bar f(q)} \frac{\partial^2 \ln \bar f(q)}{ \partial\,(\ln q)^2}
\nonumber\\
&
&+\epsilon(\eta)\,\frac{\partial\,\ln\,{\left\langle \Gamma_S(\vec k_1) \Gamma_S(\vec k_2) \right\rangle}'}{\partial \eta}
+k_1^i \,\frac{\partial \,\ln {\left\langle \Gamma_S(\vec k_1) \Gamma_S(\vec k_2) \right\rangle}'}{\partial \,k_1^i}
+k_2^i \,\frac{\partial \,\ln {\left\langle \Gamma_S(\vec k_1) \Gamma_S(\vec k_2) \right\rangle}'}{\partial \,k_2^i}
\nonumber
\\
&&+\beta_q(\eta)
\frac{\partial \,
\ln {\left\langle \Gamma_S(\vec k_1) \Gamma_S(\vec k_2) \right\rangle}'
}{\partial \ln q}\,
+\beta_n(\eta)\,n^j \,\frac{\partial\,\ln {\left\langle \Gamma_S(\vec k_1) \Gamma_S(\vec k_2) \right\rangle}' }{\partial \,n^j}\,.
\end{eqnarray}
Notice that the contributions in the first line of eq \eqref{modgenga} that depend only
on 
the  long mode (without being weighted by $\delta_\GW$)
 do not contribute to  ${\cal M}$. Therefore 
they do not modulate the short-mode two point function.

\bigskip

We now apply to the results derived  above  the definitions of $\delta_\GW$ and $\Gamma$ power spectra, eqs \eqref{defGps}, \eqref{defDps}. We find
the following expression for the modulation of the power spectrum due to a long mode:
\begin{eqnarray}
&&P_{\hat \delta_\GW}(\eta,\,k,\,q,\,\hat n,\,\vec k_L)=
\Bigg[1
+ 2\,\frac{\partial  \ln P_\zeta  }{\partial \ln k } \,\zeta_{\vec k_L}\,
+2\,  \beta_q(\eta)\,\frac{\partial \ln q}{ \partial \ln \bar f(q)} \frac{\partial^2 \ln \bar f(q)}{ \partial\,(\ln q)^2}
 \,\zeta({\vec k_L})\,
\nonumber
\\
&& \quad\quad\quad\quad\quad\quad+
\left(
\epsilon(\eta)\,\frac{\partial  \ln  \left| {\it T}_S \right|^2}{\partial \eta}\,
+\frac{\partial  \ln \left| {\it T}_S \right|^2}{\partial \ln k}\,
+\beta_n(\eta)\,\frac{\partial  \ln \left| {\it T}_S \right|^2}{\partial \ln \mu}\,
\right) \,\zeta({\vec k_L})
\Bigg]\,P_{ \delta_\GW}(\eta,\,k,\,q,\,\hat n) \,.
\nonumber\\
\label{mod2pt}
\end{eqnarray}
 All quantities inside the square parenthesis in the RHS are again evaluated at the same values of $\eta$, $\hat n$, $k$; hence  we understand this dependence. We find that 
 the power spectrum of $\delta_\GW$ is modulated by the long mode $\zeta({\vec k_L})$ through three (physically distinct) effects:
\begin{enumerate}
\item A modulation due to the scale dependence of the primordial curvature spectrum, as in Maldacena's consistency relation.
This is contained in the first line of eq \eqref{mod2pt}, second term in the RHS. (Notice that the contributions coming from derivatives of  the $1/k^3$ factor cancel out, as expected.)
\item A contribution due to the momentum-dependence of the background distribution $\bar f(q)$. This is contained in 
the first line of  eq \eqref{mod2pt}, third term in the RHS. This is a close relative of the effect pointed out by Alba and Maldacena \cite{Alba:2015cms}, although it is not exactly the same result because
we find contributions depending on second derivatives of the function $\bar f(q)$.
\item A contribution due to the time, scale, and direction dependence of the transfer function of scalar modes. This is 
 contained in 
the second line of  Eq. \eqref{mod2pt}.

\end{enumerate}

%\subsection{Modulation of the three-point correlation function }
%\label{sec_bisp}
\bigskip

\noindent
In the previous discussion 
%In Section \ref{sec_mod_2pt}
we learned how the long mode modulates the 2-point function. This effect is expected to lead to a non-vanishing squeezed limit for the 3-point function
involving the anisotropies $\delta_\GW$. Indeed, expressing a large scale limit 
of $\delta_\GW$ in terms of $\zeta$
as
% depending on $\vec k_3$ for $k_3$ small. This can be written as
\begin{equation}
  \hat \delta_\GW(\eta,\,k_3^i,\,q,\, n^i)\,=\,-\frac{\partial \ln \bar f(q)}{ \partial\,\ln q}\, {\it T}_S(\eta,\,k_3^i,\,\mu_3)\,\zeta({\vec k_3})\,,
\end{equation}
for a small $|\vec k_3|$, we can write the schematic relation (all $\delta_\GW$'s are evaluated at the same values of $\eta$, $n^i$, $q$ so we understand their dependence)
\begin{eqnarray}
\lim_{\vec k_3\to0}\,\langle \hat \delta_\GW(\vec k_1 ) \hat \delta_\GW(\vec k_2) \hat \delta_\GW(\vec k_3) \rangle
&=&-\frac{\partial \ln \bar f(q)}{ \partial\,\ln q}\, {\it T}_S(\eta,\,k_3^i,\,\mu_3)\,\left\langle \langle \hat \delta_\GW(\vec k_1 ) \hat \delta_\GW(\vec k_2)  \rangle\, \zeta({\vec k_3})  \right\rangle
\nonumber
\\
&=&-\frac{\partial \ln \bar f(q)}{ \partial\,\ln q}\, {\it T}_S(\eta,\,k_3^i,\,\mu_3)\,\left\langle \langle  \delta_\GW(\vec k_1 )  \delta_\GW(\vec k_2)  \rangle\,
\left(1+ {\cal M}\,\zeta_L\right)\, \zeta({\vec k_3}) \right\rangle
\nonumber
\\
&=&-\frac{\partial \ln \bar f(q)}{ \partial\,\ln q}\, {\it T}_S(\eta,\,k_3^i,\,\mu_3)\,{\cal M}\,\left\langle \langle  \delta_\GW(\vec k_1 )  \delta_\GW(\vec k_2)  \rangle\,\langle \zeta_L \zeta({\vec k_3}) \right\rangle\,,
\nonumber
\\
\end{eqnarray}
where in the second line we used eq \eqref{tpmod1}. This non-vanishing result gives the squeezed limit of the three-point function for $\delta_\GW$. We adopt the following definition
 \footnote{We use $P_\zeta(k_3)$ instead of $P_{\delta_\GW}(k_3)$ in the next equation, in order to simplify the overall coefficients in the equations that come next. Recall that 
 the definitions of $P_\zeta$  and $P_{\delta_\GW}$ are related by Eq. \eqref{defDps}.}
 for the non-linear
parameter $f_\NL^{\delta_\GW}$:
% that can be extracted from the squeezed limit of t
\begin{eqnarray}
\lim_{\vec k_3\to0}\,\left\langle  \delta_\GW(\vec k_1 )  \delta_\GW(\vec k_2) \delta_\GW(\vec k_3) \right\rangle\,=\,f_\NL^{\delta_\GW}\,\left(\frac{4\pi^4}{k_1^3\,k_3^3} \right)\,P_{\delta_\GW}(k_1)\,P_\zeta(k_3)\,.
\end{eqnarray}
In our case, using the previous results, we find

\begin{eqnarray}
f_\NL^{\delta_\GW}&=&-\frac{\partial \ln \bar f(q)}{ \partial\,\ln q}\, {\it T}_S(\eta,\,k_3,\,\mu_3)\,\Big[2\,\frac{\partial  \ln P_\zeta  }{\partial \ln k_1 } 
+ 2   \beta_q(\eta)\,\frac{\partial \ln q}{ \partial \ln \bar f(q)} \frac{\partial^2 \ln \bar f(q)}{ \partial\,(\ln q)^2}+
\nonumber\\
&&\hskip4cm+  
\epsilon(\eta)\,\frac{\partial  \ln  \left| {\it T}_S \right|^2}{\partial \eta}\,
+\frac{\partial  \ln \left| {\it T}_S \right|^2}{\partial \ln k_1}\,
+\beta_n(\eta)\,\frac{\partial  \ln \left| {\it T}_S \right|^2}{\partial \ln \mu_1}\,
 \Big]\,,
\nonumber
\\
\end{eqnarray}
and we can apply to this result  the very same considerations made after eq \eqref{mod2pt}.
% the end of Section \ref{sec_mod_2pt}.

\bigskip
The formula simplifies considerably in the case of pure matter domination. In this case, ${\it T}_S\,=\,3/5$, $\beta_q\,=\, {\cal H} \,\epsilon\,=\,2/5$. Then,
\begin{eqnarray} \label{fNLfor}
f_\NL^{\delta_\GW}&=&-\frac65\,\frac{\partial \ln \bar f(q)}{ \partial\,\ln q}\,
\frac{\partial  \ln P_\zeta  }{\partial \ln k_1 }  
-\frac{12}{25}\,\frac{\partial^2 \ln \bar f(q)}{ \partial\,(\ln q)^2}\,.
\end{eqnarray}
Recalling that $ \bar f(q)$ is related with the GW isotropic energy density $\Omega_\GW$ by the relation
\begin{equation}
\frac{\partial \ln \bar f}{\partial \ln q}\,=\,\frac{\partial \ln \Omega_{GW}}{\partial \ln q}-4\,,
\end{equation}
the non-linearity parameter $f_\NL^{\delta_\GW}$ can then be enhanced in proximity to large values of second derivatives of $ \Omega_\GW$ as a function of the scale $q$.

\bigskip
\begin{figure}[t!]
\begin{center}
\includegraphics[width = 0.65 \textwidth]{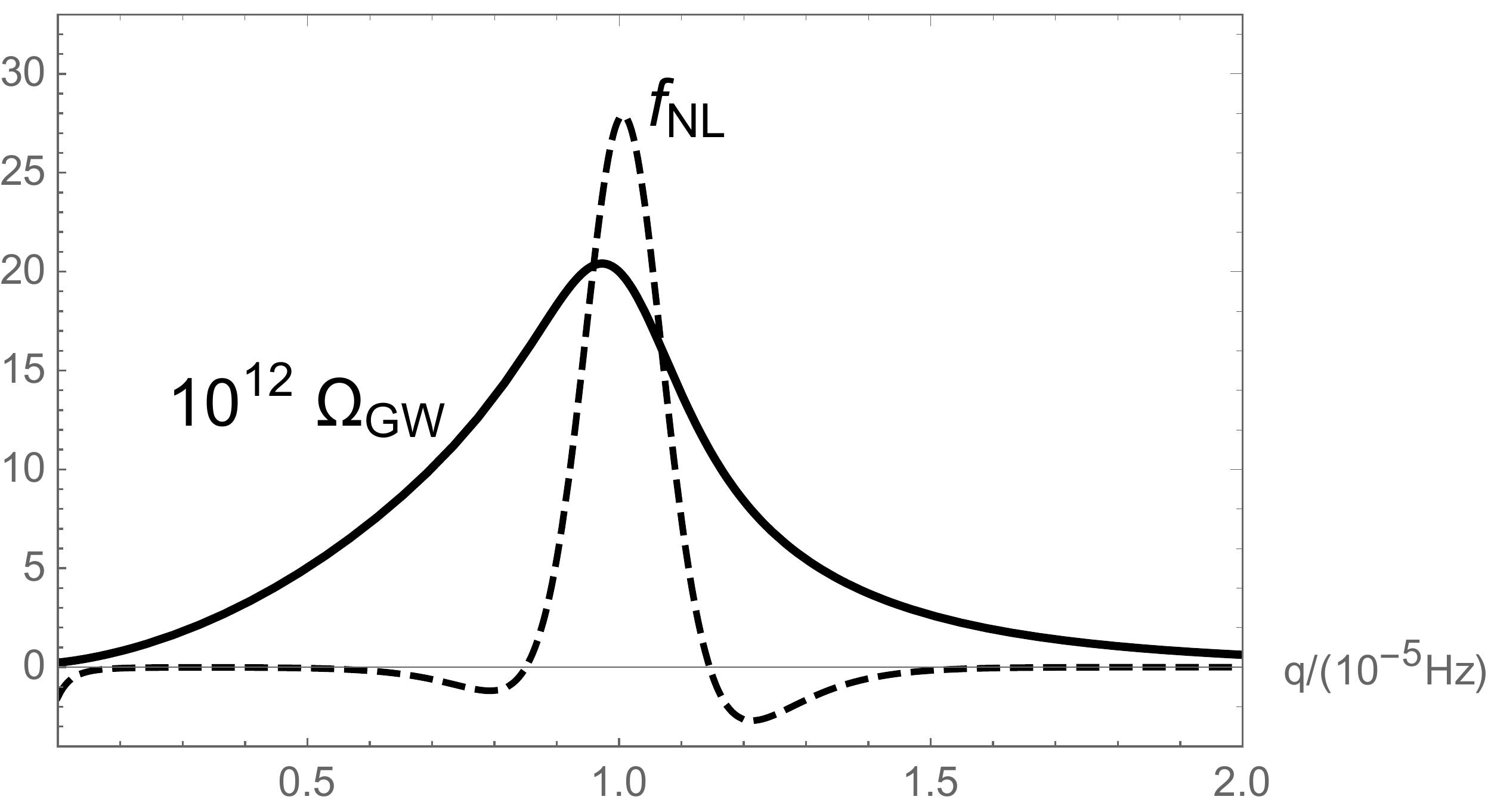}
 \caption{\it 
 Representation of the  GW spectral density $\Omega_\GW$  and of  $f_\NL^{\delta_\GW}$ for the model  given in Eqs. \eqref{toyomega},
 \eqref{toyfnl}, choosing  a scale invariant $P_\zeta$.  Notice that the magnitude of $f_\NL^{\delta_\GW}$ is  amplified 
 around the position where the spectral density changes slope. We have chosen the parameters $\alpha=2$, $\beta=5$, $\kappa_0=1/10$, $\Omega_{0}\,=\,10^{-12}$, $q_\star\,=\,10^{-5}$ Hz$^{-1}$.  }
\label{fig:art1}
\end{center} 
\end{figure}
As an illustrative toy model which demonstrates this effect, we can consider a GW spectral density
with the shape of a broken power law.
The following
parameterisation for the spectral energy changes slope at a scale
$q=q_\star$:
\begin{equation}\label{toyomega}
\Omega_\GW(q)\,=\,\frac{\Omega_{0}}{2}\left\{
\left( \frac{q}{q_\star} \right)^\alpha\,\left[ \tanh{\left[\frac{(1-q/q_\star)}{\kappa_0} \right]}+1\right]
+
\left( \frac{q}{q_\star} \right)^{-\beta}\,\left[ \tanh{\left[\frac{(q/q_\star-1)}{\kappa_0} \right]}+1\right]
%+q^{-\beta}\,\left( \tanh{\left(\frac{q-1}{ q_0} \right)}+1\right)
\right\}
\end{equation}
with $\alpha$, $\beta$ positive numbers, while the functions inside the square parenthesis represent  a regularisation  of twice the Heaviside
function (that is approached when sending $ \kappa_0\to0$). The function $\Omega_\GW$ has a large second derivative in proximity of the scale $q_\star$  where the change of slope occurs.   
The value of $f_\NL^{\delta_\GW}$ at  $q_\star$   results (for a scale invariant spectrum of $\zeta$)
\begin{equation}\label{toyfnl}
f_\NL^{\delta_\GW}\,=\,\frac{3}{25}\,\frac{\alpha+\beta}{\kappa_0}\,\left(4-(\alpha+\beta)\,\kappa_0\right)\;.
\end{equation}
Hence it can be enhanced taking small values of $\kappa_0$. See Fig \ref{fig:art1} for an illustration of this phenomenon, for a representative choice of parameters.

\section{Conclusion}
\label{sec: conclusion}

The amount of information extracted from the detection of GW signals by the LIGO-Virgo collaboration has shown the power 
of GW to study astrophysical compact object and to give relevant cosmological information on the late time universe. At the same level, the improving angular resolution of future GW detectors will allow one  to extract precious information from the detection of the stochastic background of GWs generated both from the superposition of unresolved astrophysical  sources and from cosmological sources, like inflation, phase transition or topological defects. However, high sensitivity alone will be not sufficient for discriminating among different contributions. So it becomes necessary to characterize such backgrounds using observables that can give a clear hint about the origin of the signals. As recently studied, a parity-violating SGWB, which represents a smoking gun for some cosmological signals, can be probed using ground and space-based interferometers \cite{Domcke:2019zls}. Another important tool is the directionality dependence of the SGWB. As shown for astrophysical GW, the distribution of sources implies that  the energy density  is characterized by an anisotropic contribution beyond the isotropic one. In the same way we expect that, analogously to   CMB photons, 
 also primordial GW are charaterized by anisotropies that can be generated both at the moment of production and during their propagation. 
In this paper we focused on the stochastic background of cosmological origin and we studied the anisotropies due to the production mechanism (that we encode in an initial condition term) plus those generated from the propagation of GW on the perturbed universe,  using a Boltzmann approach. We solved the Boltzmann equation for the graviton distribution function considering a FLRW metric with both scalar and tensor inhomogeneities. We showed that, contrary to CMB photons, 
at the moment of production, GWs, which are characterized by a non-thermal  spectrum, generically result in angular anisotropies that have an order one dependence on the GW frequency. We provide a criterion to evaluate whether and how much the GW anisotropies depend on frequency.  As an example, we evaluate this criterion in the case where an axion inflaton $\phi$ sources gauge fields, which in turn generates a large GW background. 

Additional anisotropies are induced by the GW propagation in the the large-scale scalar and tensor perturbations of the universe. We compute the angular power spectrum of the SGWB energy density, and, analogously to CMB photons, also the gravitons distribution function gets mainly affected by the Sachs-Wolfe effect on large scales, while the Integrated Sachs-Wolfe is subdominant. %From a comparison with SW contribution to the CMB anisotropies, we find that $C_\ell^{\rm SW} = ( 3/10)^2 {\widetilde C}_{\ell,S}$.  

We then focus on a second observable that can be a crucial tool in discriminating an astrophysical from a cosmological background, namely its departure from a Gaussian statistics. While we expect that the astrophysical background is Gaussian, due to central limit theorem, (some) cosmological backgrounds should shown  a non-Gaussian statistics. We computed the three-point function (bispectrum) of the SGWB energy density, which is not affected by de-correlation issues, both considering the effects at generation and due to propagation. We have shown that also the SGWB bispectrum carries a memory of the initial condition and that it is proportional to the non-Gaussianity of the scalar perturbations. In this sense, the SGWB can be used as a novel probe (beyond the CMB and the LSS) of the non-Gaussianity of the scalar perturbations. 

Finally we consider non-linear effects induced by long-wavelength scalar perturbations, which generate a modulation effect on the correlation functions of the short-wavelength modes. We identified  the effects of long modes with an appropriate coordinate transformation and we computed the effect of non-linearities in inducing a non-vanishing   squeezed limit of the SGWB three-point correlation function. We quantified the dependence of the squeezed bispectrum on the scale-dependence of the spectrum of primordial scalar fluctuations similar to Maldacena consistency relation, on the momentum dependence of the background SGWB distribution function, and on the time, scale, and direction dependence of the scalar transfer function.

In summary, in this paper we have approached the possibility to use CMB techniques to describe the cosmological SGWB trying to characterize it using peculiar features that we do not expect to have in the astrophysical background. Of course the detectability with interferometers of such effects is one crucial step to address and we plan to work on it on a future paper. At the same time we also plan to analyze several additional physical effects that we have neglected in this first paper, like the effects of neutrinos on the GW amplitude or a possible direct dependence of $\Gamma_I$ on ${\hat n}$, which would give distinctive signatures useful for the characterization.

\acknowledgments

N.B., D.B. and S.M. acknowledge partial financial support by ASI Grant No. 2016-24-H.0.  A.R.~is supported by the Swiss National Science Foundation (SNSF), project {\sl The Non-Gaussian Universe and Cosmological Symmetries}, project number: 200020-178787.  The work of G.T. is partially supported by STFC grant ST/P00055X/1.
%We thank...G.T. is partially supported by STFC grant ST/P00055X/1.

%\newpage

\appendix

%%%%%%%%%%%%%%%%%%%%
\section{Computation of the tensor sourced term}
\label{app:tensor-sourced} 
%%%%%%%%%%%%%%%%%%%%

In this appendix we present the steps from Eq. (\ref{Gamma-lmT-2}) to Eq. (\ref{Gamma-lmT-3}) of the main text. The first goal is to obtain an explicit expression for the integrand in Eq.  (\ref{Gamma-lmT-1}), when the integration variable $\vec{k}$ is oriented along the $z-$axis. In the $\left\{ + \times \right\}$ basis, related to the circular basis by $$e_{ij,\lambda} \equiv \frac{e_{ij,+} + i \lambda \,  e_{ij,\times}}{\sqrt{2}},$$ this orientation of $\vec{k}$ leads to 
\begin{equation}
e_{ij,+} \left( {\hat k}_z \right) = 
\frac{1}{\sqrt{2}} \left( \begin{array}{ccc} 
1 & 0 & 0 \\
0 & -1 & 0 \\ 
0 & 0 & 0 
\end{array} \right) \;\;,\;\; \quad\quad
e_{ij,\times} \left( {\hat k}_z \right) = 
\frac{1}{\sqrt{2}} \left( \begin{array}{ccc} 
0 & 1 & 0 \\
1 & 0 & 0 \\ 
0 & 0 & 0 
\end{array} \right) \,.
\end{equation} 
so that 
\begin{eqnarray}
&& \chi_{11} \left( {\hat k}_z \right) = - \chi_{22}  \left( {\hat k}_z \right) 
= \chi \left( \eta ,\, k \right) \, \frac{\xi_{-2} \left( \vec{k} \right) + \xi_2 \left( \vec{k} \right) }{2}\,, \nonumber\\ 
&& \chi_{12} \left( {\hat k}_z \right) =  \chi_{21}  \left( {\hat k}_z \right) 
 = \chi \left( \eta ,\, k \right) \, \frac{\xi_{-2} \left( \vec{k} \right) - \xi_2 \left( \vec{k} \right) }{2 i} \,. 
\end{eqnarray} 
while the other entries vanish. 

We decompose the GW direction ${\hat n}$ in a basis having ${\hat k}$ as the $z-$axis 
\begin{equation}
{\hat n} = \left( \sqrt{1 - \mu_{k,n}^2} \, \cos \phi _{k,n},\, \sqrt{1-\mu_{k,n}^2} \, \sin \phi_{k,n} ,\, \mu_{k,n} \right) \;, 
\label{n-deco-kz}
\end{equation}
In this basis  
\begin{equation}
- \frac{n^i \, n^j}{2}  \chi_{ij}' \left( \vec{k} = k \, {\hat k}_z \right) = - \frac{1-\mu_{k,n}^2}{4} \,  \chi' \left( \eta ,\, k \right) \left[ {\rm e}^{2 i \phi_{k,n}} \, \xi_2 \left( \vec{k} \right) +  {\rm e}^{-2 i \phi_{k,n}} \, \xi_{-2} \left( \vec{k} \right) \right] \,. 
\end{equation} 

Our goal is to compute 
\begin{equation}
\Gamma_{\ell m,T} =   \int \frac{d^3 k}{\left( 2 \pi \right)^3} \, {\rm e}^{i \vec{k} \cdot \vec{x}_0} \,  \int d^2 \Omega_n  \, \Gamma_T \left( \eta_0 ,\, \vec{k} ,\, \Omega_n \right) \, Y_{\ell m}^* \left( \Omega_n \right) \,,
\label{Gamma-ell-m-T-allk}
\end{equation} 
with the knowledge that, when $\vec{k}$ is decomposed according to (\ref{n-deco-kz}) (namely, with ${\hat k}$ directeed along thee $z-$axis), 
\begin{equation} 
\Gamma_T \left( \eta_0 ,\, \vec{k}  ,\, \Omega_{k,n} \right)  = - \frac{1-\mu_{k,n}^2}{4} \sum_{\lambda=\pm2} {\rm e}^{i \lambda \phi_{k,n}} \, \xi_\lambda \left( \vec{k} \right)  \int_{\eta_{\rm in}}^{\eta_0} d \eta \,  \chi' \left( \eta ,\, k \right) \, {\rm e}^{-i  \mu_k \left( \eta_0 - \eta \right) k} \,.
\label{Gamma-ell-m-T-kz}
\end{equation} 

We need to evaluate the integral (\ref{Gamma-ell-m-T-allk}) for a generic orientation of $\vec{k}$. On the other hand, the explicit expression of the integrand (\ref{Gamma-ell-m-T-kz}), holds only when $\vec{k}$ is oriented along the $z-$axis. We cope with this by rotating the integrand of the  $\int d^2 \Omega_n$ integration into a basis in which the direction ${\hat n}$ is decomposed according to 
Eq. (\ref{n-deco-kz}). 

To achieve this, we introduce the rotation matrix 
\begin{equation}
S \left( \Omega_k \right) \equiv \left( 
\begin{array}{ccc} 
\cos \theta_k \, \cos \phi_k & - \sin \phi_k & \sin \theta_k \cos \phi_k \\ 
\cos \theta_k \, \sin \phi_k &  \cos \phi_k & \sin \theta_k \sin \phi_k \\ 
- \sin \theta_k & 0 & \cos \theta_k 
\end{array} \right) \,,
\end{equation} 
in terms of which 
\begin{equation}
{\hat k} = S \left( \Omega_k \right) \, \left( \begin{array}{c} 0 \\ 0 \\ 1 \end{array} \right) \;\;,\;\; \quad\quad
\left( \begin{array}{c} 
\sin \theta_n \, \cos \phi_n \\ 
\sin \theta_n \, \sin \phi_n \\ 
\cos \phi_n \end{array} \right) =  S \left( \Omega_k \right) \, 
\left( \begin{array}{c} 
\sin \theta_{k,n} \, \cos \phi_{k,n} \\ 
\sin \theta_{k,n} \, \sin \phi_{k,n} \\ 
\cos \phi_{k,n} \end{array} \right) \,.
\end{equation} 
Under this rotation
\begin{equation}
Y_{\ell m}^* \left( \Omega_n \right) = \sum_{m'=-\ell}^\ell D_{m m'}^{(\ell)} \left( S \left( \Omega_k \right) \right) \, Y_{\ell m'}^* \left( \Omega_{k,n} \right) \;\;\;,\;\;\; d \Omega_n = d \Omega_{k,n} \,,
\label{rotation-Y}
\end{equation} 
where the Wigner rotation matrix are given by 
\begin{equation}
D^{(\ell)}_{ms}  \left( S \left( \Omega_k \right) \right) \equiv \sqrt{\frac{4 \pi}{2 \ell + 1}} \left( - 1 \right)^s \;\; _{-s}Y_{\ell m}^* \left( \Omega_k \right) \,,
\label{Wigner}
\end{equation} 
in terms of the  spin-weighted spherical harmonics 
\begin{eqnarray} 
_{-s}Y_{\ell m}^* \left( \Omega_k \right) &\equiv& \left( - 1 \right)^m \sqrt{\frac{ \left( \ell + m \right)! \left( \ell - m \right)! \left( 2 \ell + 1 \right)}{4 \pi \left( \ell + s \right)! \left( \ell - s \right)!}} \, \sin^{2 \ell} \left( \frac{\theta_k}{2} \right) \nonumber\\ 
& & \times \sum_{r=0}^{\ell - s} 
\left( \begin{array}{c} \ell - s \\ r  \end{array} \right) 
\left( \begin{array}{c} \ell + s \\ r + s - m \end{array} \right) 
\left( - 1 \right)^{\ell - r - s } \, {\rm e}^{i m \phi_k} \, \cot^{2 r + s - m } \left( \frac{\theta_k}{2} \right)\,.\nonumber\\
\end{eqnarray} 

With this relations, the equation (\ref{Gamma-ell-m-T-allk}) can be then rewritten as 
\begin{equation}
\Gamma_{\ell m,T} =  \int \frac{d^3 k}{\left( 2 \pi \right)^3} \, {\rm e}^{i \vec{k} \cdot \vec{x}_0} 
\sum_{m'=-\ell}^\ell  D_{m m'}^{(\ell)} \left( S \left( \Omega_k \right) \right)  \int d^2 \Omega_{k,n} \, 
Y_{\ell m'}^* \left( \Omega_{k,n} \right) \, \Gamma_T \left( \eta_0 ,\, \vec{k} ,\, \Omega_{k,n} \right) \,.
\label{Gamma-ell-m-T-app}
\end{equation} 
where now the innermost integrand is performed in a basis in which the ${\hat n}$ vector is decomposed according to (\ref{n-deco-kz}), 
so that the explicit expression (\ref{Gamma-ell-m-T-kz}) can  be used.

The inner integral evaluates to 
\begin{eqnarray} 
&& \!\!\!\!\!\!\!\! 
\int d^2 \Omega_{k,n} \, Y_{\ell m'}^* \left( \Omega_{k,n} \right) \, \Gamma_T \left( \eta_0 ,\, \vec{k} ,\, \Omega_{k,n} \right) = 
\int d^2 \Omega_{k,n} \, 
\sqrt{\frac{2 \ell + 1}{4 \pi} \, \frac{\left(\ell -m' \right)!}{\left( \ell + m' \right)!}} \, 
P_{\ell}^{m'} \left( \mu_{k,n} \right) \, {\rm e}^{-i m' \phi_{k,n}} 
\nonumber\\ 
&& \quad\quad\quad\quad  \quad\quad\quad\quad 
\times \left( - 1 \right) \frac{1-\mu_{k,n}^2}{4} 
\sum_{\lambda = \pm 2}  {\rm e}^{i \lambda \phi_{k,n}} \, \xi_\lambda \left( \vec{k} \right) 
\int_{\eta_{\rm in}}^{\eta_0} d \eta \,  \chi' \left( \eta ,\, k \right) \, {\rm e}^{-i  \mu_k \left( \eta_0 - \eta \right) k} \nonumber\\ 
&& \!\!\!\!\!\!\!\! 
= -\int_{\eta_{\rm in}}^{\eta_0} d \eta \, \chi' \left( \eta ,\, k \right) \int_{-1}^1 d \mu_{k,n}  \, 
\frac{1-\mu_{k,n}^2}{4} \, {\rm e}^{-i  \mu_k \left( \eta_0 - \eta \right) k} \,   P_\ell^{2} \left( \mu_{k,n} \right) 
 2 \pi \, \sqrt{\frac{2 \ell + 1}{4 \pi} \, \frac{\left(\ell -2 \right)!}{\left( \ell + 2 \right)!}} 
\sum_{\lambda = \pm 2} \delta_{m'\lambda } \, \xi_\lambda \left( \vec{k} \right)  \nonumber\\ 
&& \!\!\!\!\!\!\!\! 
=  \int_{\eta_{\rm in}}^{\eta_0} d \eta \, \chi' \left( \eta ,\, k \right) \, 
 \left( - i \right)^\ell \,  \frac{j_\ell \left( k \left( \eta_0 - \eta \right) \right)}{k^2 \left( \eta_0 - \eta \right)^2} \,
\sqrt{ 4 \pi \, \left( 2 \ell + 1 \right)}  \,  \sqrt{ \frac{\left(\ell +2 \right)!}{\left( \ell - 2 \right)!}} 
\frac{1}{4} \, \sum_{\lambda = \pm 2} \delta_{m'\lambda } \, \xi_\lambda \left( \vec{k} \right)  \,. 
\end{eqnarray} 
Inserting this into Eq. (\ref{Gamma-ell-m-T-app}), and using the relation (\ref{Wigner}) for the  Wigner elements we finally arrive to Eq.  (\ref{Gamma-lmT-3}) of the main text.

%%%%%%%%%%%%%%%%%%%%
\section{Tensor contribution to the GW bispectrum}
\label{app:tensor2B} 
%%%%%%%%%%%%%%%%%%%%

In this Appendix we present the steps from Eq. (\ref{T2B-1}) to Eq. (\ref{T2B-2})  of the main text.  We start by introducing the quantity 
${\cal F}_{\ell_1 \ell_2 \ell_3}^\lambda \left( k_1 , k_2 , k_3 \right) $ from Eq.  (2.6)  of \cite{Shiraishi:2010kd}:  
\begin{equation}
\left\langle \prod_{i=1}^3 \int d \Omega_{k_i} \xi_\lambda \left( \vec{k}_i \right) \,  _{-\lambda}Y_{\ell_i m_i}^* \left( \Omega_{k_i} \right) \right\rangle \equiv \left( 2 \pi \right)^3 {\cal F}_{\ell_1 \ell_2 \ell_3}^\lambda \left( k_1 , k_2 , k_3 \right) 
\left( \begin{array}{ccc} 
\ell_1 & \ell_2 & \ell_3 \\ 
m_1 & m_2 & m_3  
\end{array} \right) \,,
\label{G3calF}
\end{equation} 
(where we have also used Eq. (2.6)  of \cite{Shiraishi:2010kd} at the l.h.s.). This relation is inverted by Eq. (2.7)  of \cite{Shiraishi:2010kd}: 
\begin{eqnarray} 
{\cal F}_{\ell_1 \ell_2 \ell_3}^\lambda \left( k_1 , k_2 , k_3 \right) &=& \sum_{m_1,m_2,m_3} 
\left( \begin{array}{ccc} 
\ell_1 & \ell_2 & \ell_3 \\ 
m_1 & m_2 & m_3  
\end{array} \right) 
\int d \Omega_{k_1} \int d \Omega_{k_2} \int d \Omega_{k_3} \nonumber\\ 
&& _{-\lambda}Y_{\ell_1 m_1}^* \left( \Omega_{k_1} \right) 
_{-\lambda}Y_{\ell_2 m_2}^* \left( \Omega_{k_2} \right) 
_{-\lambda}Y_{\ell_3 m_3}^* \left( \Omega_{k_3} \right) \, \frac{1}{\left( 2 \pi \right)^3} \, 
\left\langle \xi_\lambda \left( \vec{k}_1 \right) \, \xi_\lambda \left( \vec{k}_2 \right) \, \xi_\lambda \left( \vec{k}_3 \right) 
\right\rangle \,. \nonumber\\ 
\end{eqnarray} 

We insert Eq. (\ref{G3calF}) in Eq. (\ref{T2B-1}) to obtain 
\begin{eqnarray} 
\left\langle \prod_{i=1}^3 \Gamma_{\ell_i m_i,T} \right\rangle &=&  {\cal G}_{\ell_1 \ell_2 \ell_3}^{m_1m_2m_3} 
\left( \begin{array}{ccc} 
\ell_1 & \ell_2 & \ell_3 \\ 
0 & 0 & 0 
\end{array} \right)^{-1} \, 
\sqrt{\frac{4 \pi}{\left( 2 \ell_1 + 1 \right) \left( 2 \ell_2 + 1 \right) \left( 2 \ell_3 + 1 \right) }} \nonumber\\ 
&&\times \left[ \prod_{i=1}^3 4 \pi \left( - i \right)^{\ell_i} \int \frac{k_i^2 \, d k_i}{\left( 2 \pi \right)^3} {\cal T}_{\ell,i}^T \left(  k_i ,\, \eta_0 ,\, \eta_{\rm in} \right) \right] \left( 2 \pi \right)^3 \sum_{\lambda = \pm 2} {\cal F}_{\ell_1 \ell_2 \ell_3}^\lambda \left( k_1 , k_2 , k_3 \right) \,.\nonumber\\ 
\end{eqnarray} 
where the relation (\ref{gaunt}) has also been used. We collect some of the factors in this expression into the combination 
\begin{eqnarray}
{\tilde {\cal F}}_{\ell_1 \ell_2 \ell_3}^\lambda \left( k_1 , k_2 , k_3 \right) \equiv 
\left( \begin{array}{ccc} 
\ell_1 & \ell_2 & \ell_3 \\ 
0 & 0 & 0 
\end{array} \right)^{-1} \, 
\sqrt{\frac{4 \pi}{\left( 2 \ell_1 + 1 \right) \left( 2 \ell_2 + 1 \right) \left( 2 \ell_3 + 1 \right) }} \, 
\left( 2 \pi \right)^3 {\cal F}_{\ell_1 \ell_2 \ell_3}^\lambda \left( k_1 , k_2 , k_3 \right)\,, \nonumber\\ 
\end{eqnarray} 
which then evaluates to the relation (\ref{cal-tilde-F}) in the main text. In terms of ${\tilde {\cal F}}$ we then recover Eq. (\ref{T2B-2})  of the main text.

%%%%%%%%%%%%%%%%%%%%
\section{Comparison with the CMB}
\label{app: comparison} 
%%%%%%%%%%%%%%%%%%%%

In the CMB case for a temperature $T \left( {\hat n} \right) = {\bar T} + \delta T \left( {\hat n} \right)$, we have 
\begin{eqnarray} 
{\bar f} \left( p  \right)  &=&  \frac{1}{{\rm e}^{\frac{p}{\bar T}} - 1} \nonumber\\ 
f \left( p , {\hat n} \right)  &=&  \frac{1}{{\rm e}^{\frac{p}{T \left( {\hat n} \right)}} - 1} 
=  \frac{1}{{\rm e}^{\frac{p}{\bar T}} - 1} + \frac{{\rm e}^{\frac{p}{{\bar T}}}}{\left( {\rm e}^{\frac{p}{{\bar T}}}-1 \right)^2} \, \frac{p}{\bar T} \, \frac{\delta T \left( {\hat n} \right)}{\bar T} = {\bar f} \left( p \right)  - p \, \frac{\partial  {\bar f} \left( p  \right) }{\partial p} \,  \frac{\delta T \left( {\hat n} \right)}{\bar T} 
\nonumber\\ 
\end{eqnarray} 
from which it follows 
\begin{equation}
\Gamma \left( {\hat n} \right) =  \frac{\delta T \left( {\hat n} \right)}{\bar T} \;\;\;,\;\;\; p \; {\rm independent} 
\end{equation} 

To connect with the description of the SGWB, we also define 
\begin{equation}
w_{\rm CMB} \left( p ,\, {\hat n} \right) = \frac{p^4 \, f \left( p  ,\, {\hat n} \right)}{\rho_{\rm crit}} \;\;\;,\;\;\; 
{\bar w}_{\rm CMB} \left( p  \right) =  \frac{p^4 \, {\bar f} \left( p   \right)}{\rho_{\rm crit}} 
\end{equation} 
so that we have the $p-$dependent quantity 
\begin{equation}
\delta_{\rm CMB} \left( p ,\, {\hat n} \right) \equiv  \frac{w_{\rm CMB} \left( p ,\, {\hat n} \right) - {\bar w}_{\rm CMB} \left( p \right) }{{\bar w}_{\rm CMB} \left( p  \right)} =  \frac{{\rm e}^{\frac{p}{{\bar T}}}}{ {\rm e}^{\frac{p}{{\bar T}}}-1 } \, \frac{p}{\bar T} \, \frac{\delta T \left( {\hat n} \right)}{\bar T} 
\end{equation} 
as well as the $p-$dependent quantity 
\begin{eqnarray}
4 -  \frac{\partial \ln \, {\bar \omega}_{\rm CMB} \left( \eta_0 ,\, p \right)}{\partial \ln \, p}  &=& 4 - \frac{\rho_{\rm crit}}{p^3 \, {\bar f} \left( p \right)} \left[ \frac{ 4 p^3 \, {\bar f} \left( p \right)}{\rho_{\rm crit}} + \frac{p^4}{\rho_{\rm crit}} \, \frac{\partial {\bar f} \left( p \right)}{\partial p} \right] \nonumber\\ 
&=& - p \left( {\rm e}^{\frac{p}{\bar T}} - 1 \right) \frac{-\frac{1}{\bar T} \, {\rm e}^{\frac{p}{\bar T}}}{\left( {\rm e}^{\frac{p}{\bar T}} -1 \right)^2}  = 
\frac{\frac{p}{\bar T} \, {\rm e}^{\frac{p}{\bar T}}}{ {\rm e}^{\frac{p}{\bar T}} -1 }  \;.
\end{eqnarray} 
So the ratio 
\begin{equation}
\frac{\delta_{\rm CMB} \left( p ,\, {\hat n} \right) }{4 -  \frac{\partial \ln \, {\bar \omega}_{\rm CMB} \left( \eta_0 ,\, p \right)}{\partial \ln \, p} } = \frac{\delta T \left( {\hat n} \right) }{\bar T} 
\end{equation}
is indeed $p-$independent.


\end{document}